\documentclass[sigconf, nonacm]{acmart}

\newcommand\vldbdoi{XX.XX/XXX.XX}
\newcommand\vldbpages{XXX-XXX}
\newcommand\vldbvolume{14}
\newcommand\vldbissue{1}
\newcommand\vldbyear{2020}
\newcommand\vldbauthors{\authors}
\newcommand\vldbtitle{\shorttitle} 
\newcommand\vldbavailabilityurl{https://github.com/xianghongxu/COOOL}
\newcommand\vldbpagestyle{plain} 
\usepackage{multirow}
\usepackage{balance}
\usepackage{bbold}

\def\BibTeX{{\rm B\kern-.05em{\sc i\kern-.025em b}\kern-.08emT\kern-.1667em\lower.7ex\hbox{E}\kern-.125emX}}

\usepackage{libertine}

\usepackage{booktabs} 
\usepackage{mathtools}
\usepackage{subcaption}
\usepackage{caption}

\usepackage[ruled,vlined,linesnumbered]{algorithm2e}
\usepackage[noend]{algpseudocode}
\usepackage{makecell}
\usepackage{mathrsfs} 
\usepackage{amsmath}
\usepackage{amsbsy}
\usepackage{placeins}

\usepackage{amsmath,amssymb,amsfonts} 

\DeclareMathOperator*{\argmax}{arg\,max}

\usepackage{textcomp}
\usepackage{xcolor}

\usepackage{booktabs} 
\usepackage{mathtools}
\usepackage{subcaption}
\usepackage{enumitem}
\usepackage{import}
\usepackage{graphicx}
\graphicspath{ {./images/} }

\begin{document}
\title{COOOL: A Learning-To-Rank Approach for SQL Hint Recommendations}

\author{Xianghong Xu}
\affiliation{%
  \institution{ByteDance Inc.}
  \institution{Tsinghua University}
  \city{Beijing}
  \country{China}
}
\email{xxh20@mails.tsinghua.edu.cn}

\author{Zhibing Zhao}
\affiliation{%
  \institution{ByteDance Inc.}
  \city{Seattle}
  \country{USA}
}
\email{zhibing.zhao@bytedance.com}

\author{Tieying Zhang}
\affiliation{%
  \institution{ByteDance Inc.}
  \city{San Jose}
  \country{USA}
}
\email{tieying.zhang@bytedance.com}

\author{Rong Kang}
\affiliation{%
  \institution{ByteDance Inc.}
  \city{Beijing}
  \country{China}
}
\email{kangrong.cn@bytedance.com}

\author{Luming Sun}
\affiliation{%
  \institution{ByteDance Inc.}
  \city{Beijing}
  \country{China}
}
\email{sunluming@bytedance.com}

\author{Jianjun Chen}
\affiliation{%
  \institution{ByteDance Inc.}
  \city{San Jose}
  \country{USA}
}
\email{jianjun.chen@bytedance.com}

\begin{abstract}
Query optimization is a pivotal part of every database management system (DBMS) since it determines the efficiency of query execution. Numerous works have introduced Machine Learning (ML) techniques to cost modeling, cardinality estimation, and end-to-end learned optimizer, but few of them are proven practical due to long training time, lack of interpretability, and integration cost. A recent study provides a practical method to optimize queries by recommending per-query hints but it suffers from two inherited problems. First, it follows the regression framework to predict the absolute latency of each query plan, which is very challenging because the latencies of query plans for a certain query may span multiple orders of magnitude. Second, it requires training a model for each dataset, which restricts the application of the trained models in practice. 
In this paper, we propose \textsf{COOOL} to predict \underline{C}ost \underline{O}rders of query plans to c\underline{OO}perate with DBMS by \underline{L}earning-To-Rank. Instead of estimating absolute costs, \textsf{COOOL} uses ranking-based approaches to compute relative ranking scores of the costs of query plans. We show that \textsf{COOOL} is theoretically valid to distinguish query plans with different latencies.
We implement \textsf{COOOL} on PostgreSQL, and extensive experiments on join-order-benchmark and TPC-H data demonstrate that \textsf{COOOL} outperforms PostgreSQL and state-of-the-art methods on single-dataset tasks as well as a unified model for multiple-dataset tasks. Our experiments also shed some light on why \textsf{COOOL} outperforms regression approaches from the representation learning perspective, which may guide future research. 

\end{abstract}

\maketitle

\pagestyle{\vldbpagestyle}
\begingroup\small\noindent\raggedright\textbf{PVLDB Reference Format:}\\
\vldbauthors. \vldbtitle. PVLDB, \vldbvolume(\vldbissue): \vldbpages, \vldbyear.\\
\href{https://doi.org/\vldbdoi}{doi:\vldbdoi}
\endgroup
\begingroup
\renewcommand\thefootnote{}\footnote{\noindent
This work is licensed under the Creative Commons BY-NC-ND 4.0 International License. Visit \url{https://creativecommons.org/licenses/by-nc-nd/4.0/} to view a copy of this license. For any use beyond those covered by this license, obtain permission by emailing \href{mailto:info@vldb.org}{info@vldb.org}. Copyright is held by the owner/author(s). Publication rights licensed to the VLDB Endowment. \\
\raggedright Proceedings of the VLDB Endowment, Vol. \vldbvolume, No. \vldbissue\ %
ISSN 2150-8097. \\
\href{https://doi.org/\vldbdoi}{doi:\vldbdoi} \\
}\addtocounter{footnote}{-1}\endgroup

\ifdefempty{\vldbavailabilityurl}{}{
\vspace{.3cm}
\begingroup\small\noindent\raggedright\textbf{PVLDB Artifact Availability:}\\
The source code, data, and/or other artifacts have been made available at \url{\vldbavailabilityurl}.
\endgroup
}

\section{Introduction}
\label{sec:intro}

Query optimization is vital to the performance of every database management system (DBMS). A SQL query typically has many candidate plans that are equivalent in terms of final output but differ a lot in execution latency. The goal of query optimization is to select the best candidate plan with the lowest latency for each query from a vast search space with sufficient accuracy.

Query optimization has been studied for decades \cite{oldwork_selinger1979access} and is still an active research field \cite{aidbsurvey_tsesmelis2022database}. Various ML-based research lines have been proposed: cost modeling, cardinality estimation, end-to-end query optimization, etc., among which the most practical approach is Bao~\cite{marcus2022bao}. Bao is a query optimization system leveraging tree convolutional neural networks \cite{tcnn_mou2016convolutional} and Thompson sampling \cite{sampling_thompson1933likelihood} to recommend SQL hints. Bao has made a remarkable improvement in the practicality of end-to-end query optimization, but it suffers from two problems inherited from previous models \cite{neo,liu2015cardinality,e2ecost_19_sunji,naru_yang2019deep,yang2020neurocard}. First, Bao follows the same underlying assumption as the previous works, i.e., the model needs to predict the exact cost of each plan to select the plan with the minimum cost. It first estimates the absolute cost of each candidate plan under a regression framework and then selects the one with the minimum estimate. While accurately estimating the cost is sometimes desirable, it is very challenging for existing models \cite{akdere2012learning,wu2013predicting,ventura2021expand}. As a result, the model may make inaccurate predictions and select query plans with high latencies. 
Second, Bao trains a model for each dataset respectively and evaluates the model on the corresponding dataset. The datasets vary among DBMS instances and it is costly to train and maintain individual models for every dataset. Therefore, the generalizability of the model is desirable in real-world scenarios.

To address these, we propose \textsf{COOOL} to estimate \underline{C}ost \underline{O}rders of query plans to c\underline{OO}perate with DBMS leveraging \underline{L}earning-To-Rank (LTR) techniques. \textsf{COOOL} is designed on top of an existing DBMS to recommend query-specific hints to facilitate \textit{practical} use. To obtain a unified model that can optimize queries from different datasets, we employ a data/schema agnostic TCNN as the query plan cost ranking scorer to predict relative cost orders of candidate plans. We make similar assumptions to Bao to inherit its advantages: we assume a finite set of hint sets are predefined and all hint sets result in semantically equivalent query plans. 

In contrast to regression methods to predict costs, LTR approaches learn relative scores for different hint sets. LTR approaches have advantages over regression approaches from previous end-to-end query optimization works~\cite{neo,marcus2022bao,balsa_SIGMOD22} for the following reasons. The plan execution latency ranges from several milliseconds to thousands of seconds. And a little difference in the tree structures of two semantically equivalent plans for a query may lead to a large difference in execution latencies. These facts bring difficulties in regression model training and inference because the objective function of the regression approach is to minimize the $L_2$ error. The squared error formula is sensitive to anomalous large or small latencies, which may make the model performance unstable~\cite{balsa_SIGMOD22}. Despite normalization methods can alleviate this problem by mapping the original latency distribution to a fixed interval (e.g, [0,1]), they may lead to a latency distribution distortion because the relative latency gap among query plans cannot be fully reflected. Accurately estimating the latency of every query plan is not necessary for end-to-end query optimization since the optimizer only needs to select one of them to execute. A good prediction on the orders of the latencies of query plans is sufficient for query optimization, which is exactly the objective of LTR.

LTR is a supervised learning framework to train models for ranking tasks. There are primarily three categories of LTR approaches: pointwise, pairwise, and listwise. Most pointwise methods are the same as regression. Pairwise methods concern the relative order between two items, and listwise methods focus on the order of items in the entire list. The three approaches can be applied to most existing models, e.g., neural networks, and the only difference lies in the loss function. In this work, we leverage the widely used tree convolutional neural networks (TCNN)~\cite{tcnn_mou2016convolutional,marcus2022bao,balsa_SIGMOD22} as the underlying model. To train an LTR model, one can use the orders of latencies as labels, which has the potential to be more robust than regression models against the large range of orders of magnitude in query plan latencies. By transforming the absolute cost estimation problem into relative cost order prediction, \textsf{COOOL} can utilize LTR techniques to train the TCNN so that the output tells which plan is the best. Specifically, we respectively study the performance of pairwise and listwise strategies on SQL hints recommendation for query optimization, so \textsf{COOOL} has two implementations: \textsf{COOOL}-pair and \textsf{COOOL}-list. By this means, we may also make it possible to train a unified model to improve query plans from different datasets.

Our experiments show that \textsf{COOOL} can consistently improve query plans in various settings, achieving as large as 6.73$\times$ speedup over PostgreSQL. Moreover, we investigate the regression framework and ranking strategies from the perspective of representation learning \cite{representationLearningSurvey_zhang2018network}. We show that the model trained by the regression approach has a dimensional collapse \cite{dimensionalCollapse} in the plan embedding space. Whereas \textsf{COOOL} does not have a dimensional collapse using the same embedding method. The dimensional collapse will restrict the ML methods to building a unified model because the collapsed dimensions may be different in different datasets.

Furthermore, Bao requires more effort to implement because it is fully integrated into PostgreSQL, while we build on top of PostgreSQL, which makes it easy for \textsf{COOOL} to migrate to other DBMSs.

To summarize, we make the following contributions:
\begin{itemize}
    \item We propose \textsf{COOOL}, a learned model that predicts \textbf{C}ost \textbf{O}rders of the query plans to c\textbf{OO}perate with DBMS by \textbf{L}TR techniques, to recommend better SQL hints for query optimization. To our best knowledge, \textsf{COOOL} is the first end-to-end query optimization method that maintains a unified model to optimize queries from different datasets. 
    \item We theoretically show that \textsf{COOOL} can distinguish query plans with different latencies when optimizing the loss functions, and verify that \textsf{COOOL} is superior to regression approaches from the representation learning perspective.
    \item Comprehensive experiments on join-order-benchmark and TPC-H show that \textsf{COOOL} can outperform PostgreSQL and state-of-the-art methods on multiple dimensions of evaluation criteria. 
\end{itemize}

\section{Preliminaries}
\subsection{Task Definition and Formalization}
Let $Q$ denote the set of queries and $\mathcal{H}=\{HS_1,HS_2,\ldots,HS_n\}$ be the set of $n$ hint sets. Each hint set $HS_i\in\mathcal{H}$ contains only boolean flag query hints (e.g., enable hash join, disable index scan). For any query $q\in Q$ and $i\in\{1, 2, \ldots, n\}$, the traditional optimizer $Opt$ can generate the corresponding plan tree $t^q_i$ with the hint set $HS_i$. 
\begin{align}
    t^q_i = Opt(q,HS_i).
    \label{eq:optimizerexplain}
\end{align}

Let $T^q=\{t^q_1, t^q_2, \ldots, t^q_n\}$ be the set of candidate plans of query $q$. The query plans in $T^q$ are semantically equivalent, but may have different execution latencies. 

A model $M$ is a function that takes the candidate plan tree as input and produces a score for the plan tree.
\begin{equation}\label{eq:sqi}
s^q_i = M(q, t_i^q; \vec\theta),
\end{equation}
where $\vec\theta$ is the parameter of the model to be trained. The query execution engine then selects the plan with the highest predicted ranking score in the scenario of LTR.

\begin{align}
    \hat{HS}^q = HS_{\argmax\limits_i s^q_i},
    \label{eq:selectHintset}
\end{align}
where $\hat{HS}^q$ is the hint set with the maximum score, corresponding to the minimum estimated cost.

\subsection{Learning-To-Rank (LTR)}\label{sec:Pre-LTR}

In the context of LTR, $\mathcal{H}$ is the set of items, and the goal is to recommend the best item from $\mathcal{H}$ to any $q\in Q$. More specifically, LTR is to define a loss function on $s^q_i$'s so that the underlying model $M$ can be properly trained to predict the orders of query plans. Throughout this paper, ``$t^q_{i_1}\succ t^q_{i_2}$" means the plan $t^q_{i_1}$ is superior to (has a lower latency than) $t^q_{i_2}$. Given a query $q\in Q$, let $\sigma_q = t^q_{i_1}\succ t^q_{i_2}\succ\ldots\succ t^q_{i_n}$ denote the total order of query plans w.r.t. their latencies, where $t^q_{i_1}$ has the lowest latency, $t^q_{i_2}$ has the second lowest latency, and $t^q_{i_n}$ has the highest latency.

Before introducing our pairwise loss function, we will first introduce a classic model, the Plackett-Luce model (PL)~\cite{Plackett75:Analysis,Luce59:Individual}, based on which we selected our pairwise method. PL is one of the most popular models for discrete choices, which was later used as a listwise loss function in information retrieval~\cite{Liu09:Learning} and softmax function in classification tasks~\cite{Gao17:Properties}. In the context of SQL hint recommendations, we provide a definition of PL as follows. Given any $q\in Q$, the probability of $\sigma_q = t^q_{i_1}\succ t^q_{i_2}\succ\ldots\succ t^q_{i_n}$ is
\begin{equation}\label{eq:pl}
\Pr\nolimits_{\text{PL}}(t^q_{i_1}\succ t^q_{i_2}\succ\ldots\succ t^q_{i_n}; \vec\theta)=\prod^n_{j=1}\frac {\exp(s^q_{i_j})} {\sum^n_{m=j}\exp(s^q_{i_m})},
\end{equation}
where $s^q_i$'s are functions of the parameter $\vec\theta$ defined in Equation \eqref{eq:sqi}. The rankings of multiple queries are assumed to be independent. Therefore, the probability of multiple rankings is simply the product of the probability of each individual ranking. The marginal probability of any pairwise comparison $t^q_{i}\succ t^q_{j}$ is 
\begin{equation}\label{eq:plpair}
\Pr\nolimits_{\text{PL}}(t^q_{i}\succ t^q_{j}; \vec\theta) = \frac {\exp(s^q_{i})} {\exp(s^q_{i}) + \exp(s^q_{j})}.
\end{equation}

\subsubsection{Listwise Loss Function}

Given the training data $\{\sigma_q | \forall q\in Q\}$, the listwise loss function is simply the negative log likelihoood function:
\begin{equation}\label{eq:listloss}
\mathcal{L}_\text{list}(\vec\theta) = -\sum_{q\in Q}\ln\Pr\nolimits_{\text{PL}}(\sigma_q; \vec\theta),
\end{equation}
where $\Pr\nolimits_{\text{PL}}(\sigma_q; \vec\theta)$ is defined in Equation \eqref{eq:pl}. This listwise loss function also coincides the listMLE loss by \cite{listmle_xia2008listwise}. \cite{listmle_xia2008listwise} proved that this loss function is consistent, which means as the size of the dataset goes to infinity, the learned ranking converges to the optimal one.

\subsubsection{Pairwise Loss Function}

To apply a pairwise loss function, the full rankings $\sigma_q$ for all $q\in Q$ need to be converted to pairwise comparisons. This process is called rank-breaking~\cite{Azari13:Generalized}. A rank-breaking method defines how the full rankings should be converted to pairwise comparisons. Basic breakings include full breaking, adjacent breaking, and others~\cite{Azari13:Generalized}. Full breaking means extracting all pairwise comparisons from a ranking, and adjacent breaking means extracting only adjacent pairwise comparisons. For example, given a ranking $t_1\succ t_2\succ t_3$, full breaking converts it to ($t_1\succ t_2$, $t_1\succ t_3$, $t_2\succ t_3$) while adjacent breaking converts it to ($t_1\succ t_2$, $t_2\succ t_3$). Though adjacent breaking is simple and plausible, \cite{Azari13:Generalized} proved that adjacent breaking can lead to inconsistent parameter estimation, where means even if the model is trained using an infinite amount of data, it may not make unbiased predictions. Other breakings are more complicated and beyond the scope of this paper.

Let $P=\{\pi_1,\pi_2, \ldots, \pi_m\}$ be the dataset of pairwise comparisons, where for any $j\in\{1, 2, \ldots, m\}$, $\pi_j$ has the form of $t^{q_j}_{i_1}\succ t^{q_j}_{i_2}$. Here $q^j$ denotes the corresponding query for $\pi_j$. For different $j$ values, $q_j$ may refer to the same query since different pairwise comparisons can be extracted from the same query. The objective function is

\begin{equation}\label{eq:pairloss}
\mathcal{L}_\text{pair}(\vec\theta) = -\sum^m_{j=1}\ln\Pr\nolimits_{\text{PL}}(\pi_j; \vec\theta),
\end{equation}
where $\Pr\nolimits_{\text{PL}}(p_j; \vec\theta)$ is defined in Equation \eqref{eq:plpair}. The model parameter can be estimated my maximizing $L(\vec\theta)$.

This is equivalent to maximizing the composite marginal log likelihood based on Equation \eqref{eq:plpair}. \cite{Khetan16:Data} and \cite{Zhao18:Composite} proved that $\vec\theta$ can be efficiently estimated by  under a correct rank breaking method. And full breaking, which extracts all pairwise comparisons from the full rankings, is one of the correct breaking methods.

\section{COOOL Architeture}
\begin{figure}
    \centering
    \includegraphics[width=\linewidth]{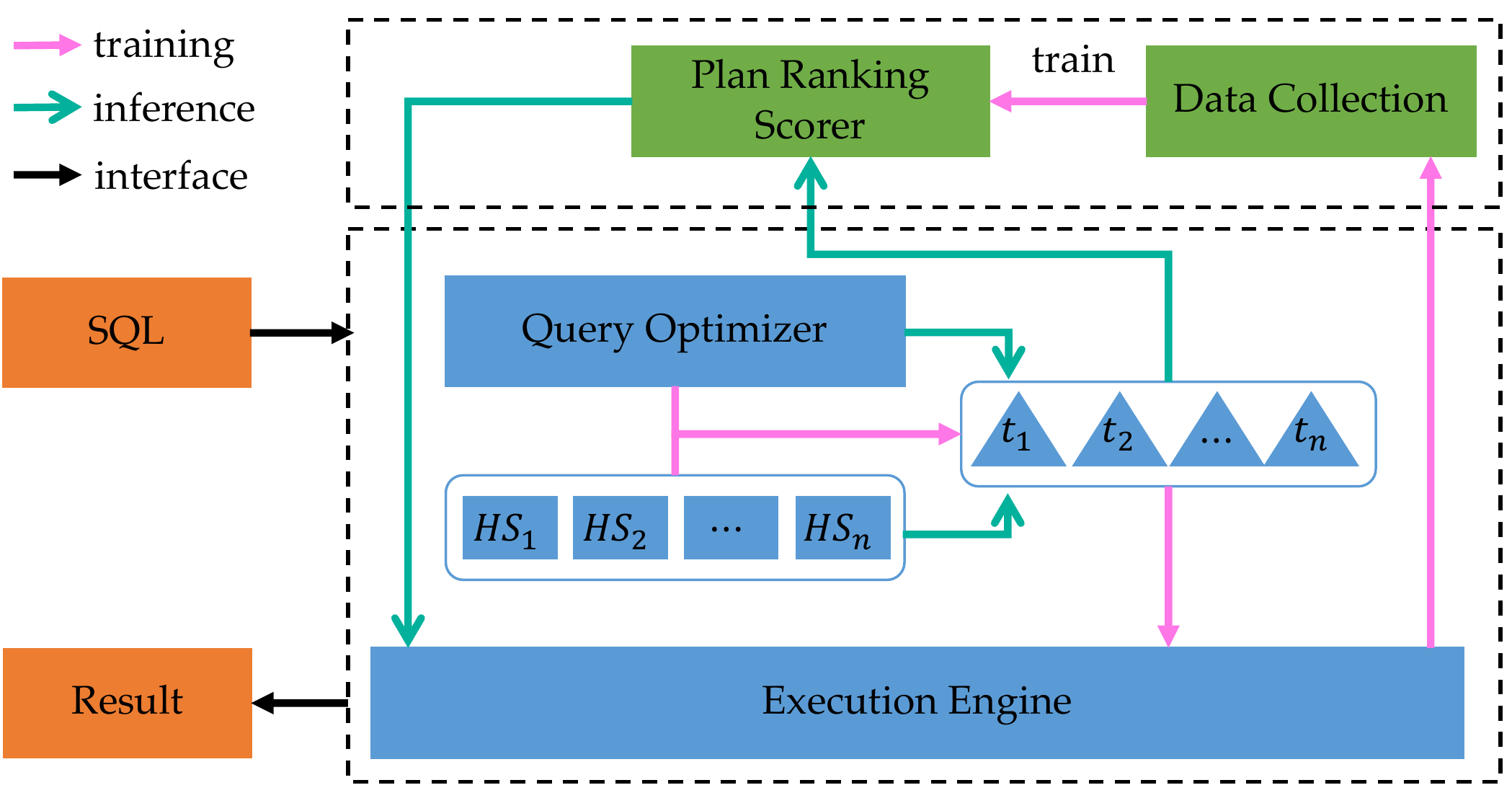}
    \caption{A brief view of the \textsf{COOOL} pipeline.}
    \label{fig:model_architecture}
\end{figure}

The core component of \textsf{COOOL} is a neural plan ranking scorer, which estimates the relative ranking scores of candidate plans. At the training stage, we execute query plans and collect their performance metrics as training data. Then we utilize the pairwise or listwise approach to implement training the ranking scorer. At the inference stage, when a user submits a query, the traditional optimizer will generate $n$ query plans by utilizing the corresponding hint sets. Next, the scorer will compute the relative ranking score of each plan and recommend the optimal one to the execution engine. The data flow pipeline of \textsf{COOOL} is shown in Figure \ref{fig:model_architecture}.

\paragraph{Data Collection} Our approach is trained in a standard supervised learning paradigm. For the given queries, we generate $n$ query plans that correspond to the hint sets $\mathcal{H}$ for each query by the underlying traditional optimizer. Then we sent them to the execution engine and record the observed execution performance of each query plan. The collected data is used to train the neural ranking scorer, and the training stage is separate from the DBMS.

\paragraph{Cost Order Estimation} To execute the optimal plan with minimum estimated latency, we need to recommend its corresponding hint set to the DBMS. The scorer takes a plan tree as input and outputs the ranking score of the plan, the estimated latency orders can be acquired by sorting the scores of all candidate plans. Specifically, we first transform the nodes of the input plan tree into vectors, then feed the vector tree into a plan embedding model constructed by a TCNN. Finally, we feed the plan embedding into a multilayer perceptron (MLP) to compute the estimated score of the plan. At the training stage, we use the collected plan and latency data to train the model. So it can estimate relative orders of plans by latency and cooperate with DBMS to improve query plans at the inference stage.

\paragraph{Assumptions and Comparisons} We assume that applying each hint set to the given query will generate semantically equivalent plans. Besides, the hints are applied to the entire query rather than the partial plan. Though allowing fine-grained hints (e.g., allows nested loop joins between specific tables, others not.) are available, it will bring an exponential candidate plan search space, which significantly increases training and inference overhead. We make the same assumption as Bao due to practicality.

Bao requires more effort to implement because it is fully integrated into PostgreSQL, while we build on top of PostgreSQL, which makes it easy for \textsf{COOOL} to migrate to other DBMSs. Moreover, we take a step forward to maintain a unified model to optimize queries from different datasets, which has not been investigated in the previous end-to-end query optimization studies.

\section{COOOL for Hint Recommendations}
First, we introduce our cost order estimation model that takes query plans as input and computes the relative ranking scores. Next, we take advantage of the pairwise and listwise approaches to formalize the training loop.  Finally, we elaborate theoretical analysis of how ranking losses work to select the best plan.

\subsection{Cost Order Estimation}
Each hint set corresponds to a query plan tree, so recommending the optimal hint set for given queries is to select the plan tree with maximum cost ranking score, as shown in Equation \eqref{eq:selectHintset}. Similar to \cite{neo,marcus2022bao,balsa_SIGMOD22}, we use a TCNN to obtain plan embeddings and leverage an MLP to compute the ranking scores.

\begin{figure*}
    \centering
    \includegraphics[width=\linewidth]{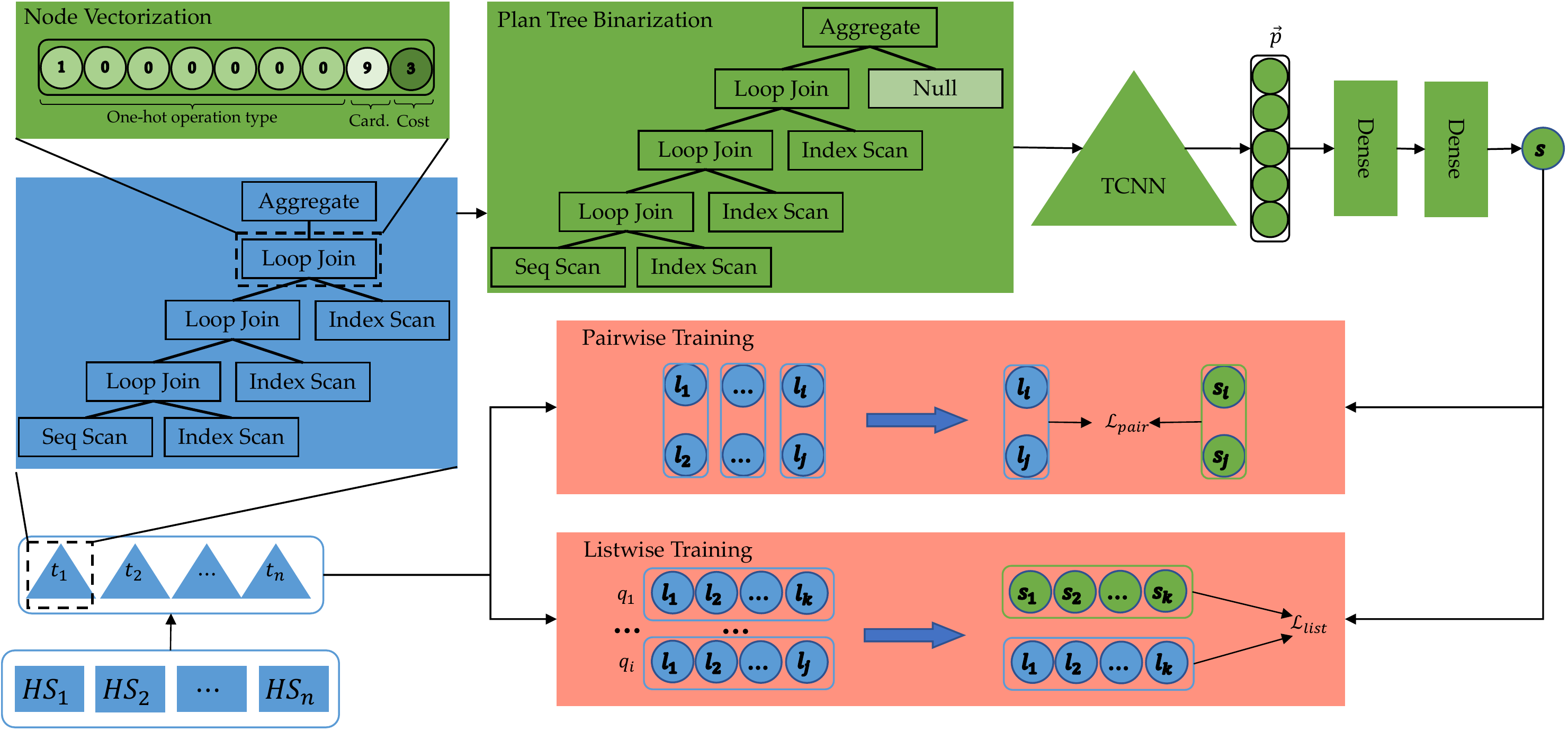}
    \caption{Cost order estimation for tree convolutional neural network using pairwise and listwise LTR techniques.}
    \label{fig:model}
\end{figure*}

\paragraph{Plan Tree Vectorization} We can use the \textsf{EXPLAIN} command provided by the underlying optimizer to obtain the plan tree text, as shown in Equation \eqref{eq:optimizerexplain}. First, we need to transform each node in the plan tree into a vector. The same as \cite{marcus2022bao}, we use a data/schema agnostic encoding scheme, which solely contains a one-hot encoding of operator types, and cardinality and cost provided by the underlying traditional optimizer. Then, we transform the original tree into a binary tree to facilitate tree convolution operations.

\paragraph{One-hot Node Operation Type Encoding.} We summarize the types of all operations (nested loop, hash join, merge join, seq scan, index scan, index only scan, and bitmap index scan) in the plan trees and number these seven operations. Then we create a vector for each node with the number of types of bits, and we set the bit of the type corresponding to each node to the high bit. For instance, $E_o(v)$ is the one-hot encoding of node $v$ and $E_o(v)[i]=1$ indicates node $v$ is the $i$-th operation type and the rest elements in $E_o(v)$ are 0. Though one-hot node type encoding is simple, it is capable to extract structural information in plan trees.

\paragraph{Tree Nodes Vectorization.} Apart from the operation information, each node can contain the cost and cardinality. It is applicable to acquire cost and cardinality from multiple traditional optimizers and learned models, but we only use two values obtained from the underlying traditional optimizer for simplicity. Therefore, the node encoding is the concatenation of operation type encoding, cardinality, and cost, i.e., $E(v)=\mathsf{Concat}(E_o(v),\mathsf{Cost}(v),\mathsf{Card}(v))$, where $\mathsf{Cost}(v)$ and $\mathsf{Card}(v)$ are the cost and cardinality estimated by the traditional optimizer, respectively. We apply the node encoding method to the nodes in the plan tree to obtain a vectorized plan tree.

\paragraph{Tree Structure Binarization.} Some nodes in a plan tree may have only one child, e.g., nodes for aggregation and sorting operations. To facilitate tree convolution operations, we transform the non-binary trees into binary trees by adding a pseudo-child node \textsf{Null} to each node with only one child, and the costs and cardinalities of pseudo-child nodes are 0. Then the original plan tree $t$ can be transformed into vectorized tree $p$.

\subsubsection{TCNN Plan Embedding} TCNN was proposed in \cite{tcnn_mou2016convolutional} to treat tree  structure data in programming language processing. TCNN was first introduced in plan representation in \cite{neo}, and it was well-established in \cite{marcus2022bao,balsa_SIGMOD22}. We will briefly introduce how to represent a plan using TCNN in this section, refer to \cite{tcnn_mou2016convolutional,neo} for more technical details.

During the execution of the original plan tree in execution engines, the computation of one node relies on the results of its child nodes. Based on this fact, the plan embedding method should reflect the recursive properties to obtain a proper inductive bias \cite{inductivebias_mitchell1980need}. To be consistent with plan execution, the model is naturally required to simultaneously capture the features of a node and its child nodes. Specifically, denote $l(v)$ and $r(v)$ are the left and right child nodes of node $v$, respectively. The statistic cost/cardinality information in vector $E(v)$ is closely related to $E(l(v))$ and $E(r(v))$. Tree convolution can naturally address this request.

Tree convolution is similar to image convolution, it has binary tree structure filters to capture local features. We take a tree convolution filer as an example, there are three weight vectors in the filter, i.e., $w,w_l,w_r$. Applying tree convolution to the current node $E(v)$ can acquire the new representation:
\begin{align}
    E(v)^\prime = \sigma(E(v)\odot w+ E(l(v))\odot w_l + E(r(v))\odot w_r),
\end{align}
where $\sigma$ is a non-linear activation function, $\odot$ is a dot product operation. The new representation of node $v$ contains its child nodes' information. By stacking tree convolution operations, the latent representations will have a larger receptive field than just being able to interact with child nodes. By this means, the model is able to capture high-level features of a long chain of plan execution for representing one node. The output of tree convolution operations is a tree with the same structure as the input, so we employ a dynamic pooling method to aggregate the latent representations of all nodes to represent the query plan. To sum up, we can obtain plan embedding by $\vec{p}=\mathrm{TCNN}(p)$, where $\vec{p}\in\mathbb{R}^h$ is the vector of plan representation, $h$ is the size of plan tree embedding space.

\subsubsection{Ranking Score Computation} Finally, we can leverage plan embedding to compute the relative ranking score. We use a simple MLP to take the plan embedding vector as input and output a scalar as the ranking score $s$, i.e., $s=\mathrm{MLP}(\vec{p})$. Stacking fully connected layers and non-linear activation functions in the MLP can enhance the representation ability, and this practice has been widely adopted in query optimizations \cite{neo,balsa_SIGMOD22}. We add a hidden layer and an activation function for simplicity. The cost orders can be obtained by sorting the ranking scores of all candidate plans.

\subsection{Learning-To-Rank Training Loop}

The training loop consists of three parts: data collection and deduplication, label mapping and pairwise data extraction, and model training and evaluation.

\subsubsection{Data Collection and Deduplication} Let $Q_{\mathrm{train}}$ denote the set of training queries. We generate $n$ plans for each query $q\in Q_{\mathrm{train}}$ by the traditional optimizer, as shown in Equation \eqref{eq:optimizerexplain}. Then we sent the plans to the execution engine, and record each data point $(\mathrm{query}=q,\mathrm{plan}=t,\mathrm{latency}=l)$. There are duplicate query plans because for a given query, different hints may result in the same query plan. We remove the duplicate query plans for pairwise and listwise training loops.

\subsubsection{Label Mapping and Pairwise Data Extraction} Because a lower latency indicates a better plan, we use the reciprocal of the latency of each query plan as the label to reverse the orders of query plans. Any other mapping function that reverses the orders works equivalently because only the orders matter. 

After the label mapping, we get a list of query plans ordered by the reciprocals of their latencies for each query. For each list of $n^q$ query plans ($n^q\le n$ after deduplication), we extract all $n^q\choose 2$ pairwise comparisons to get the pairwise data $P=\{\pi_1,\pi_2, \ldots, \pi_m\}$, where for any $j\in\{1, 2, \ldots, m\}$, $\pi_j=t^{q_j}_{i_1}\succ t^{q_j}_{i_2}$.

\subsubsection{Model Training and Evaluation} We compute the model parameter $\vec\theta$ by maximizing the pairwise or listwise log-likelihood function. Specifically, for the listwise approach, the model parameter $\vec\theta$ is computed by minimizing the listwise loss defined in Equation \eqref{eq:listloss}. For the pairwise approach, $\vec\theta$ is computed by minimizing the pairwise loss defined in Equation \eqref{eq:pairloss}. Refer to Section \ref{sec:Pre-LTR} for more details.

During the inference stage, we use the learned model to compute the score for each candidate plan and sort the scores to obtain the corresponding orders, then we select the estimated best plan for each query to execute.

\subsection{Theoretical Analysis}

In this section, we briefly analyze how \textsf{COOOL} learns from the order of latencies of different query plans. To show that, we consider the query plans $\{t_1, t_2, \ldots, t_n\}$ with the corresponding latencies $\{l_1, l_2, \ldots, l_n\}$ for a given query $q\in Q$. And our goal is to select the best one using the model. At the training stage, they have different initial scores $\{s_1, s_2, \ldots, s_n\}$. Without loss of generality, we assume $l_1>l_2>\ldots >l_n.$

\subsubsection{Pairwise Approach}

We start with the pairwise approach because it is easier and more intuitive than the listwise approach. In the pairwise approach, each line of data consists of two query plans for a certain query $q\in Q$, denoted by $t^q_{i_1}$ and $t^q_{i_2}$, respectively. Without loss of generality, we assume their corresponding latencies $l_{i_1}>l_{i_2}$. And let $s_{i_1}$ and $s_{i_2}$ denote the model outputs for the two query plans, and define $\delta=s_{i_2}-s_{i_1}$. For simplicity, we focus on this pair of query plans and let $\mathcal{L}^{q, i_1, i_2}_{pair}$ denote the loss on this pair of query plans. We have
\begin{equation*}
    \begin{aligned}
        \mathcal{L}^{q, i_1, i_2}_{pair}&=-\ln\frac {\exp(s_{i_2})} {\exp(s_{i_1})+\exp(s_{i_2})}\\
        &=-\ln\frac {\exp(s_{i_1}+\delta)} {\exp(s_{i_1})+\exp(s_{i_1}+\delta)}\\
        &=-\ln\frac {\exp(\delta)} {1+\exp(\delta)}\\
        &=-\delta+\ln(1+\exp(\delta))
    \end{aligned}
\end{equation*}
for $s_{i_1}=0$ because the loss only depends on $\delta$.

The partial derivative of $\mathcal{L}^{q, i_1, i_2}_{pair}$ w.r.t. $\delta$ as
\begin{equation}\label{eq:dLddelta_pair}
    \begin{aligned}
        \frac{\partial \mathcal{L}^{q, i_1, i_2}_{pair}}{\partial \delta_i}&=-1+\frac {\exp(\delta)} {1 + \exp(\delta)}<0.
    \end{aligned}
\end{equation}

This means an increase in $\delta$ leads to a decrease in the loss function $\mathcal{L}^{q, i_1, i_2}_{pair}$. $\delta$ tends to go up while the loss function $\mathcal{L}^{q, i_1, i_2}_{pair}$ is being minimized during the training process, which is desired.

\subsubsection{Listwise Approach}

We are interested in the difference between the model outputs of adjacent query plans, i.e., $\delta_i = s_{i+1} - s_i$ for all $i\in\{1, 2, \ldots, n-1\}$. For convenience, we define $\delta_0=0$. Then for all $i=\{1, 2, \ldots, n\}$, we have $$s_i = s_1 + \sum^{i-1}_{j=0}\delta_j.$$

The loss function of the listwise approach can be written as:
\begin{equation*}
    \begin{aligned}
        \mathcal{L}_{list}&=-\ln\prod^n_{j=1}\frac {\exp(s_{n-j+1})} {\sum^{n-j+1}_{m=1}\exp(s_{m})} \\
        &=-\sum^n_{j=1}\ln\frac{\exp(s_1 + \sum^{n-j}_{k=0}\delta_k)}{\sum^{n-j+1}_{m=1}\exp(s_1 + \sum^{m-1}_{k=0}\delta_k)}\\
        &=-\sum^n_{j=1}\ln\frac{\exp(\sum^{n-j}_{k=0}\delta_k)}{\sum^{n-j+1}_{m=1}\exp(\sum^{m-1}_{k=0}\delta_k)}\\
        &=-\sum^n_{j=1}\left(\sum^{n-j}_{k=0}\delta_k - \ln(\sum^{n-j+1}_{m=1}\exp(\sum^{m-1}_{k=0}\delta_k))\right).
    \end{aligned}
\end{equation*}

In this equation, the first equality is obtained by substituting Equation \eqref{eq:pl} in Equation \eqref{eq:listloss}, and the second equality is obtained by substituting $s_i$ with $s_1 + \sum^{i-1}_{j=0}\delta_i$. The third equality is obtained by dividing both the numerator and denominator by $\exp(s_1)$ since $\exp(s_1) > 0$. And the last equality is due to the property of the $\ln()$ function.

Now we compute the partial derivative of $\mathcal{L}_{list}$ with respect to $\delta_i$ for any $i\in\{1, 2, \ldots, n-1\}$:
\begin{equation}\label{eq:dLddelta_list}
    \begin{aligned}
        \frac{\partial \mathcal{L}_{list}}{\partial \delta_i}&=-\sum^{n-i}_{j=1}\left(1 - \frac {\sum^{n-j+1}_{m=i+1}\exp(\sum^{m-1}_{k=0}\delta_k)} {\sum^{n-j+1}_{m=1}\exp(\sum^{m-1}_{k=0}\delta_k)}\right)<0.
    \end{aligned}
\end{equation}

The inequality holds because when $1\le i\le n-j$, we have $$\sum^{n-j+1}_{m=i+1}\exp(\sum^{m-1}_{k=0}\delta_k) < \sum^{n-j+1}_{m=1}\exp(\sum^{m-1}_{k=0}\delta_k).$$
This means for each $i\in\{1, 2, \ldots, n\}$, an increase in $\delta_i$ leads to a decrease in the loss function $\mathcal{L}_{list}$. $\delta_i$ tends to go up while the loss function $\mathcal{L}_{list}$ is being minimized during the training process, which is desired.

To summarize, by minimizing $\mathcal{L}_{list}$ or $\mathcal{L}_{pair}$, the differences between the ranking scores of different query plans tend to increase, which means that our approaches are able to distinguish the best plans from others.

\section{Experiments}
\label{Experiments}
To comprehensively evaluate the performance of \textsf{COOOL}, we conduct experiments on three scenarios: \textit{single instance}, \textit{workload transfer}, and \textit{maintaining one model} (a unified model). The first scenario is common in machine learning for query optimization, which refers to learning a model for each dataset. The others are rarely concerned in related studies but crucial for machine learning model deployment in DBMS, especially the last scenario implies training a model to improve query plans from different datasets.

Therefore, we conduct extensive experiments to primarily answer the following Research Questions (RQs):
\begin{itemize}
    \item {\em RQ1: can \textsf{COOOL} achieve the best performance compared to the baseline methods in terms of total query execution latency and improving slow queries?}
    \item {\em RQ2: is it able to directly transfer a schema agnostic model to another dataset?}
    \item {\em RQ3: can the proposed methods improve query plans from different datasets by maintaining a unified model?}
    \item {\em RQ4: can the experiments provide some insight on why are \textsf{COOOL} methods better than Bao?}
\end{itemize}

\subsection{Experimental Setup}

\paragraph{Datasets and Workloads}
We use two widely used open-source datasets (IMDB, TPC-H) and the corresponding workloads (JOB, TPC-H). To make a fair comparison and facilitate reproduction, we do not modify the original queries.

\begin{itemize}
    \item \textbf{Join Order Benchmark (JOB)}. JOB \cite{JOB_leis2015good} contains 113 analytical queries, which are designed to stress test query optimizers over the Internet Movie Data Base (IMDB) collected from the real world. These queries from 33 templates involve complex joins (ranging from 3 to 16 joins, averaging 8 joins per query). 
    \item \textbf{TPC-H}. TPC-H \cite{tpch_poess2000new} is a standard analytical benchmark that data and queries are generated from uniform distributions. We use a scale factor of 10. There are 22 templates in TPC-H, we omit templates \#2 and \#19 because some nodes in their plan trees have over two child nodes, which makes tree convolution operation unable to handle. For each template, we generate 10 queries by the official TPC-H query generation program\footnote{\url{https://www.tpc.org/tpc\_documents\_current\_versions/current\_specifications5.asp}}.
\end{itemize}

\paragraph{Baseline Methods.} We compare our proposed model with traditional and state-of-the-art methods as follows.
\begin{itemize}
    \item PostgreSQL: We use the optimizer of PostgreSQL (version 12.5) itself with default settings.
    \item Bao: We substantially optimize the Bao source code\footnote{\url{https://github.com/learnedsystems/baoforpostgresql}} as follows. First, we used all 48 hint sets in Bao paper, rather than the 5 hint sets in the open-sourced code. Second, because we use the standard benchmarks without modifying queries, we train Bao on all past sufficiently explored execution experiences of the training set.
\end{itemize}

\paragraph{Model Implementation}
We use a three-layer TCNN and the number of channels are respectively set as \{256, 128, 64\}, the size of the plan embedding space $h$ is 64, and the hidden size of MLP is set as 32. The activation function is Leaky ReLU \cite{leakyrelu_xu2015empirical}, the optimizer is Adam \cite{adam_kingma2014adam} with an initial learning rate of 0.001, we apply an early stopping mechanism in 10 epochs on the training loss, we save the model that performed best on the validation set and report the results on the test set. 

\paragraph{Device and Configuration}
We use a virtual machine with 16 GB of RAM, and an 8-core Xeon(R) Platinum 8260 2.4GHz CPU. To achieve excellent DBMS performance, we configure PostgreSQL knobs suggested by PGTune\footnote{\url{https://pgtune.leopard.in.ua/\#/}}: 4 GB of shared buffers, 12 GB of effective cache size, etc. We implement our model in Pytorch 1.12.0 and use an NVIDIA V100 GPU.

\paragraph{Evaluation Protocol}
We repeat each experiment 10 times and exclude the best and the worst test results except for the unified model scenario, then we report the average performance of the rest runs. The total execution latency of the test set is defined as the sum of per-query execution latency. The validation set is 10\% of the training set, except for TPC-H ``repeat" settings, which use 20\% of the training set.

\subsection{Single Instance Experiments (RQ1)}

\begin{table*}[t]
    \centering
    \caption{Total query execution latency speedups on single datasets over PostgreSQL. The best performance on each workload is in boldface.}
    \begin{tabular}{c|cccccccc}
    \toprule
        & \multicolumn{4}{c}{JOB} & \multicolumn{4}{c}{TPC-H} \\
        & adhoc-rand & adhoc-slow & repeat-rand & repeat-slow & adhoc-rand & adhoc-slow & repeat-rand & repeat-slow \\
    \midrule
        Bao & 1.07 & 0.91 & 3.02 & 1.37 & 5.36 & 1.17 & 5.28 & 4.73\\
        \textsf{COOOL}-list & \textbf{1.35} & \textbf{1.46} & {3.01} & \textbf{1.57} & \textbf{6.09} &  3.63 & \textbf{5.33} & {5.55}\\
        \textsf{COOOL}-pair & 1.30 & 1.36 & \textbf{3.47} & 1.56 & 3.86 &\textbf{3.86} & {5.28} & \textbf{5.56}\\
    \bottomrule
    \end{tabular}
    \label{tab:single}
\end{table*}

In this section, we focus on the performance of different optimization methods in the \textit{single instance} scenario. According to the evaluation details in previous works \cite{marcus2022bao,balsa_SIGMOD22}, there are primarily two kinds of evaluation settings in the scenario of single instance optimization: 1) randomly split the standard benchmark into train/test sets. 2) Take the standard benchmark as the test set, and augment queries by randomly replacing predicates so that the training data can cover all the templates in the test set. These two evaluation aspects are not comprehensive, so we adapt them as follows while using the standard benchmark without adding extra queries.

\paragraph{Evaluation Criteria for Model Performance} Instead of selecting the test data uniformly at random, we evaluate the model on multiple dimensions of evaluation criteria so we have several ways to test data selection. First, we consider two real-world scenarios: ad hoc and repetitive. 

In the \textbf{ad hoc} scenario, the queries are from templates that are not included in the training data. More concretely, we first select several templates from the dataset and use all queries from the selected templates as the test data. This corresponds to the real-world scenario where the model needs to recommend a hint set for a query that is very different from previous queries. We use queries from seven templates in the JOB dataset and four templates in the TPC-H dataset respectively as test sets. 

In the repetitive scenario, the queries in the test data are ``similar" to the queries in the training data, but not the same. Here ``similar" means the queries are from the same templates as in the training data. In practice, we take one or multiple queries from each template as the test set and keep the remaining queries in the training set. For the JOB dataset, we take one query from each template and for the TPC-H dataset, we take two queries from each template as test data. This scenario is denoted as \textbf{repeat}. For queries from the same template, we take their average execution latency to represent the corresponding template's latency.

For each of the two scenarios, we comprehensively evaluate the performance of the model under the random and the tail latency evaluation protocols, so we respectively conduct random templates/queries selection and slow templates/queries selection, abbreviated as \textbf{rand} and \textbf{slow} respectively.

\begin{figure}[t]
     \centering
     \begin{subfigure}[b]{0.45\textwidth}
         \centering
         \includegraphics[width=\textwidth]{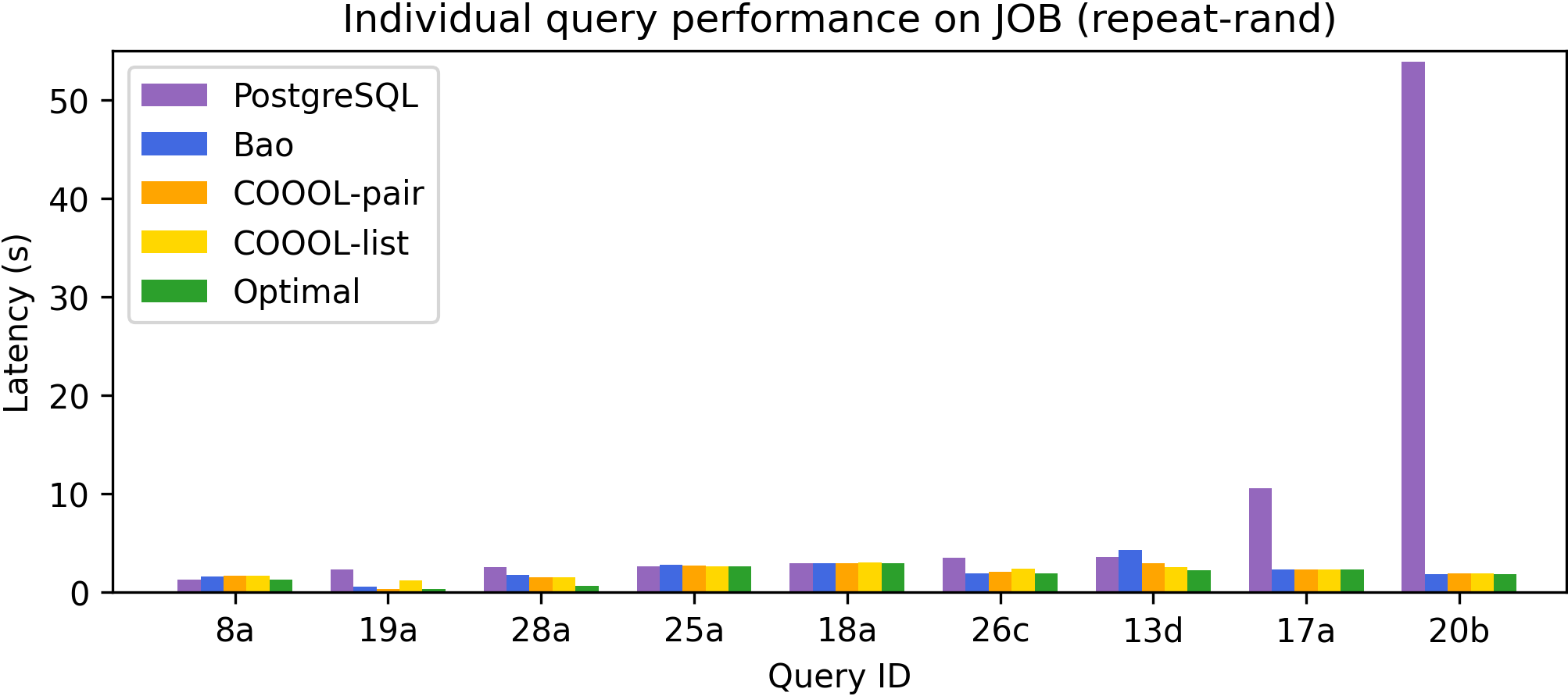}
         \caption{JOB repeat-rand}
     \end{subfigure}
     \hfill
     \begin{subfigure}[b]{0.45\textwidth}
         \centering
         \includegraphics[width=\textwidth]{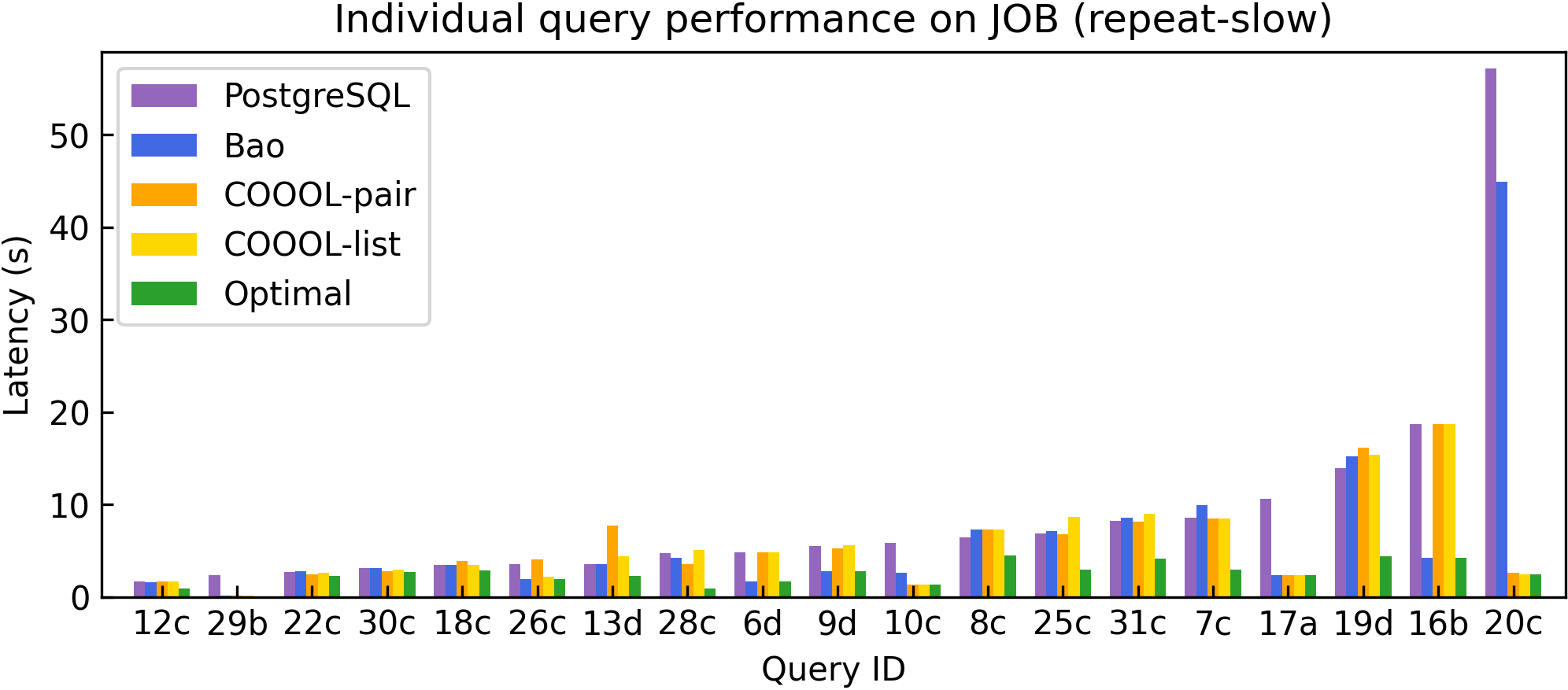}
         \caption{JOB repeat-slow}\label{fig:jobrepeatslow-single}
     \end{subfigure}
     \hfill
     \begin{subfigure}[b]{0.45\textwidth}
         \centering
         \includegraphics[width=\textwidth]{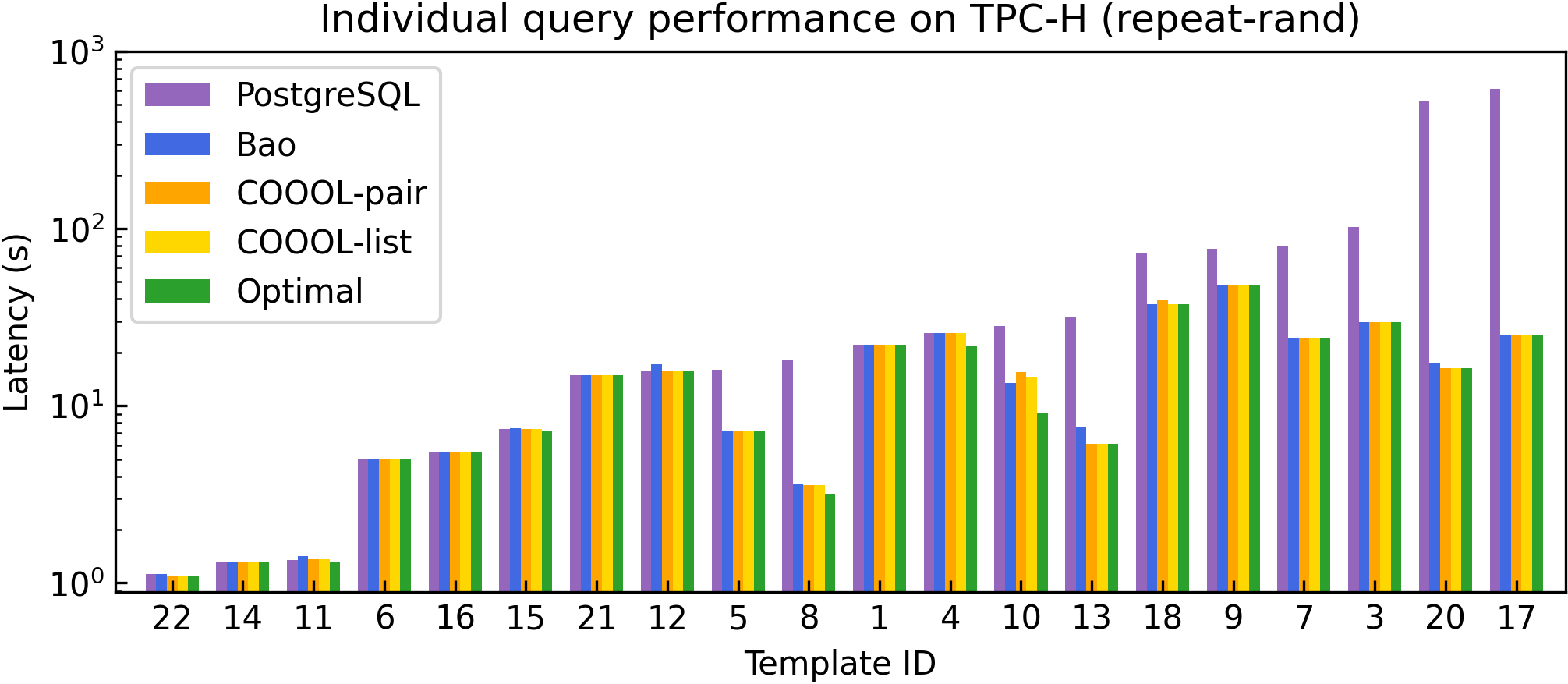}
         \caption{TPC-H repeat-rand}
     \end{subfigure}
     \hfill
     \begin{subfigure}[b]{0.45\textwidth}
         \centering
         \includegraphics[width=\textwidth]{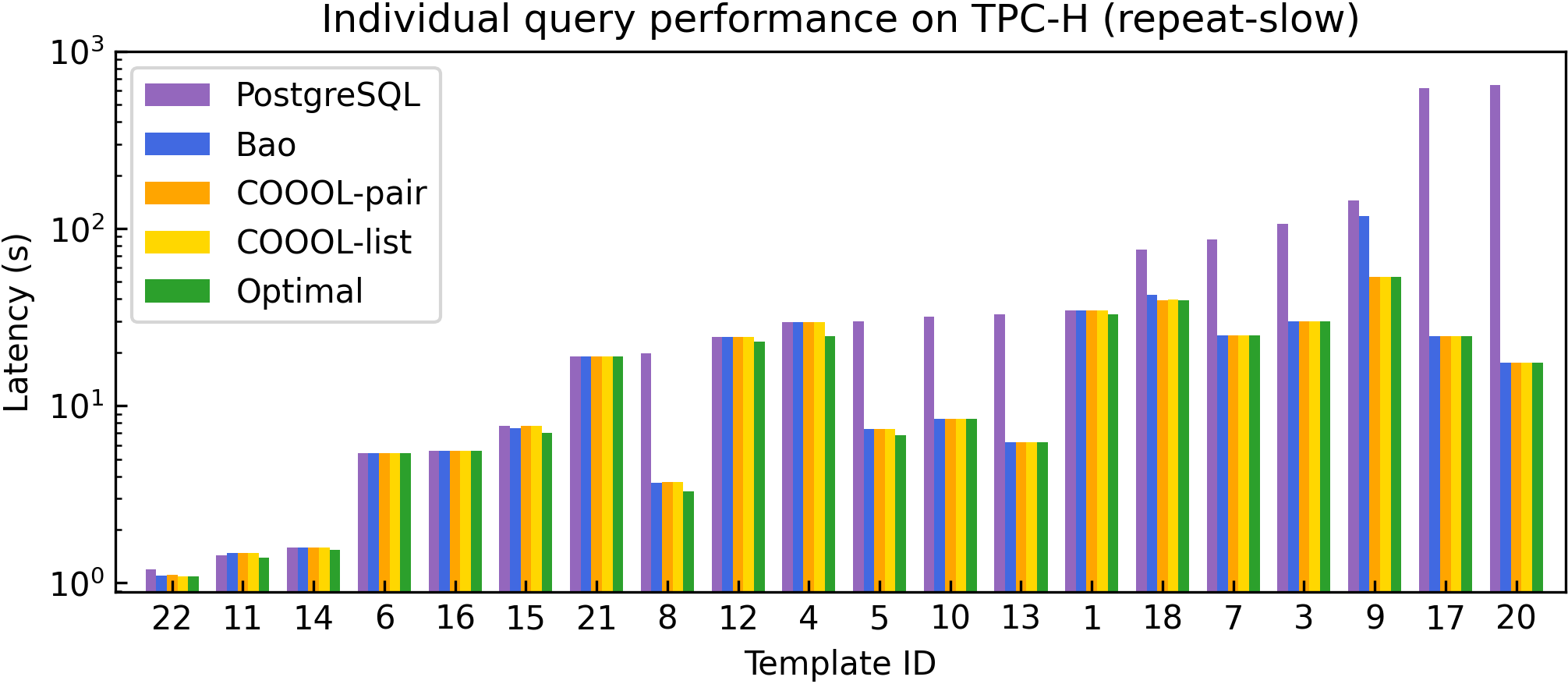}
         \caption{TPC-H repeat-slow}\label{fig:tpchrepeatslow-single}
     \end{subfigure}
     \hfill
    \caption{Individual query performance of the models in the single instance scenario.}
    \label{fig:single_repeat}
\end{figure}

Overall results on single instance experiments are summarized in Table~\ref{tab:single}. The individual query performance in ``repeat" settings is shown in Figure \ref{fig:single_repeat}, where we depict queries with an execution latency greater than 1s on PostgreSQL to facilitate observation, and ``Optimal" represents the lowest latency under the given $\mathcal{H}$.

\paragraph{Observations}

In Table~\ref{tab:single}, there are respectively four settings for each of the two workloads and we observe that:
\begin{itemize}
    \item The listwise approach (\textsf{COOOL}-list) achieves the best performance in most settings, and beats Bao by large margins in almost all settings except for ``repeat-rand" on JOB, where \textsf{COOOL} is slightly slower than Bao.
    \item The pairwise approach (\textsf{COOOL}-pair) also outperforms Bao in almost all settings except for ``adhoc-rand" on TPC-H. It achieves the best performance on three settings (``repeat-rand" on JOB, ``adhoc-slow" and ``repeat-slow" on TPC-H). In most settings, the performances of \textsf{COOOL}-pair and \textsf{COOOL}-list are similar.
    \item Bao does not have the best performance under any of these settings. For ``adhoc-slow" on JOB, Bao even has a total execution latency \textit{regression}, which means the running speed is lower than the PostgreSQL optimizer itself.
\end{itemize}

Combined with Figure \ref{fig:single_repeat}, we have the following observations:
\begin{itemize}
    \item Bao is significantly worse than \textsf{COOOL} on some queries under ``slow" settings (20c in ``repeat-slow" on JOB and template 9 in ``repeat-slow" on TPC-H.). ``Slow" settings are obviously more challenging for ML models than ``rand" settings and \textsf{COOOL} has clear advantages over Bao. 
    \item Both \textsf{COOOL} and Bao are close to optimal on ``repeat-rand" scenarios for JOB and TPC-H data. 
    \item It happens that Bao or \textsf{COOOL} are not as good as PostgreSQL on a small amount of queries, which is normal for ML models. Nevertheless, performance regressions occur less frequently on \textsf{COOOL} models than Bao models in most settings, and \textsf{COOOL}-pair has the lowest number of query regressions for these settings. See Table~\ref{tab:single-instance-regression} for details.
\end{itemize}

\begin{table}[t]
    \centering
    \caption{Number of regressions for Bao and \textsf{COOOL} compared with PostgreSQL (single instance)}
    \begin{tabular}{cccc}
    \toprule
        Setting & Bao & \textsf{COOOL}-list & \textsf{COOOL}-pair \\
    \midrule
        JOB repeat-rand & 24 & 17 & \textbf{13}\\
        JOB repeat-slow & 17 & 11 & \textbf{8}\\
        TPC-H repeat-rand & 3 & \textbf{1} & \textbf{1}\\
        TPC-H repeat-slow & \textbf{1} & \textbf{1} & \textbf{1}\\
    \bottomrule
    \end{tabular}
    \label{tab:single-instance-regression}
\end{table}

To summarize, we observe that \textsf{COOOL} has advantages and competitiveness over Bao in speeding up total query execution, alleviating individual query performance regression, and optimizing slow queries. Besides, although \textsf{COOOL}-list is better than \textsf{COOOL}-pair in total query execution latency speedups, it is not as good in avoiding individual query regression.

\subsection{Workload Transfer Investigation (RQ2)}
\label{sec:workload-transfer}

It is well-known that an instance-optimized model may have a poor performance on another workload because it does not learn patterns from the unseen one.
Even though most ML models are schema specific which makes establishing a learned model on another workload not applicable, little experimental research has been conducted for schema agnostic models. Here we present an intuitive view of directly transferring an instance-optimized model to another workload. Specifically, we train a model on a \textit{source} workload and then test its performance on another workload, namely \textit{target} workload (\textit{source}$\to$\textit{target}).

We conduct experiments on training on JOB and transfer the model to TPC-H (JOB$\to$TPC-H). And we train a model on TPC-H and investigate its performance on JOB (TPC-H$\to$JOB). In this directly transferring investigation, we also explore them in ``adhoc" and ``repeat" scenarios and under ``rand" and ``slow" settings, the same as instance-optimized experiments. To make an intuitive comparison, we train the model on the source workload's training set and show the performance on the target workload's test set.

The overall query execution speedups are shown in Table \ref{tab:cross}, and we have the following observations.

\begin{itemize}
    \item The difference of the performances of models is unstable compared with the corresponding instance-optimized models, and most settings have a performance decline. Our experiments show that  directly applying a model trained on one workload to another workload cannot obtain good performances even if the model is schema agnostic. 
    \item Compared with instance-optimized models, most models learned on TPC-H perform worse on JOB. By contrast, there is a performance improvement on ``TPC-H adhoc" settings especially on ``TPC-H adhoc-slow". We may conclude that the data in JOB may benefit TPC-H and especially improve the performance of slow queries. We will provide a deeper analysis from a representation learning perspective in Section \ref{sec:RepresentationLearning}.
    \item We show the plan tree statistics for nodes and depth of the two workloads in Table \ref{tab:plan-statistic}. On the one hand, it indicates that JOB is more complicated than TPC-H, so directly transferring a model learned on JOB may optimize queries from TPC-H. On the other hand, it suggests that data distributions in different datasets affect the performance of machine learning models in workload transfer. Because the experiment results on ``TPCH repeat" settings demonstrate that a model learned on the templates from JOB cannot achieve the same performance as that learned on the template from TPC-H.  
\end{itemize}

Based on the observations, introducing JOB data has a better performance than the model trained on TPC-H ``adhoc" settings. We can conclude that for a given workload, using the training data from the same workload may not obtain satisfactory performance. Therefore, how to utilize data from other workloads is an emerging problem for query optimization models.

\begin{table}[t]
    \centering
    \caption{Overall plan tree statistics of the two workloads.}
    \begin{tabular}{ccccccc}
    \toprule
        Workload & Max Nodes & Avg. Nodes &Max Depth &Avg. Depth \\
    \midrule
        JOB   & 72 & 23.6 & 36 & 12.0 \\
        TPC-H & 35 & 14.3 & 20 & 9.6  \\
    \bottomrule
    \end{tabular}
    \label{tab:plan-statistic}
\end{table}

\begin{table*}[t]
    \centering
    \caption{Total query execution latency speedups on the target workload  over PostgreSQL of direct transfer, where $\uparrow$ indicates an increase in performance compared to the corresponding instance-optimized model.}
    \begin{tabular}{c|cccccccc}
    \toprule
        & \multicolumn{4}{c}{TPC-H $\rightarrow$ JOB} & \multicolumn{4}{c}{JOB $\rightarrow$ TPC-H} \\
        & adhoc-rand & adhoc-slow & repeat-rand & repeat-slow & adhoc-rand & adhoc-slow & repeat-rand & repeat-slow \\
    \midrule
        Bao & \textbf{1.07}& \textbf{1.19}$\uparrow$ & \textbf{1.07} & 0.69 & \textbf{6.35}$\uparrow$ & 1.56$\uparrow$ & \textbf{1.64} & 1.42\\
        \textsf{COOOL}-list & 0.97 & 0.93 & {0.92} & 0.82 & 0.85 & \textbf{4.70}$\uparrow$ & \textbf{1.64} & 1.70\\
        \textsf{COOOL}-pair & 0.96 & 1.18 & 0.89 & \textbf{0.86} & 5.90$\uparrow$ & 4.60$\uparrow$ & 1.48 & \textbf{1.82}\\

    \bottomrule
    \end{tabular}
    \label{tab:cross}
\end{table*}

\begin{figure}[t]
     \centering
     \begin{subfigure}[b]{0.45\textwidth}
         \centering
         \includegraphics[width=\textwidth]{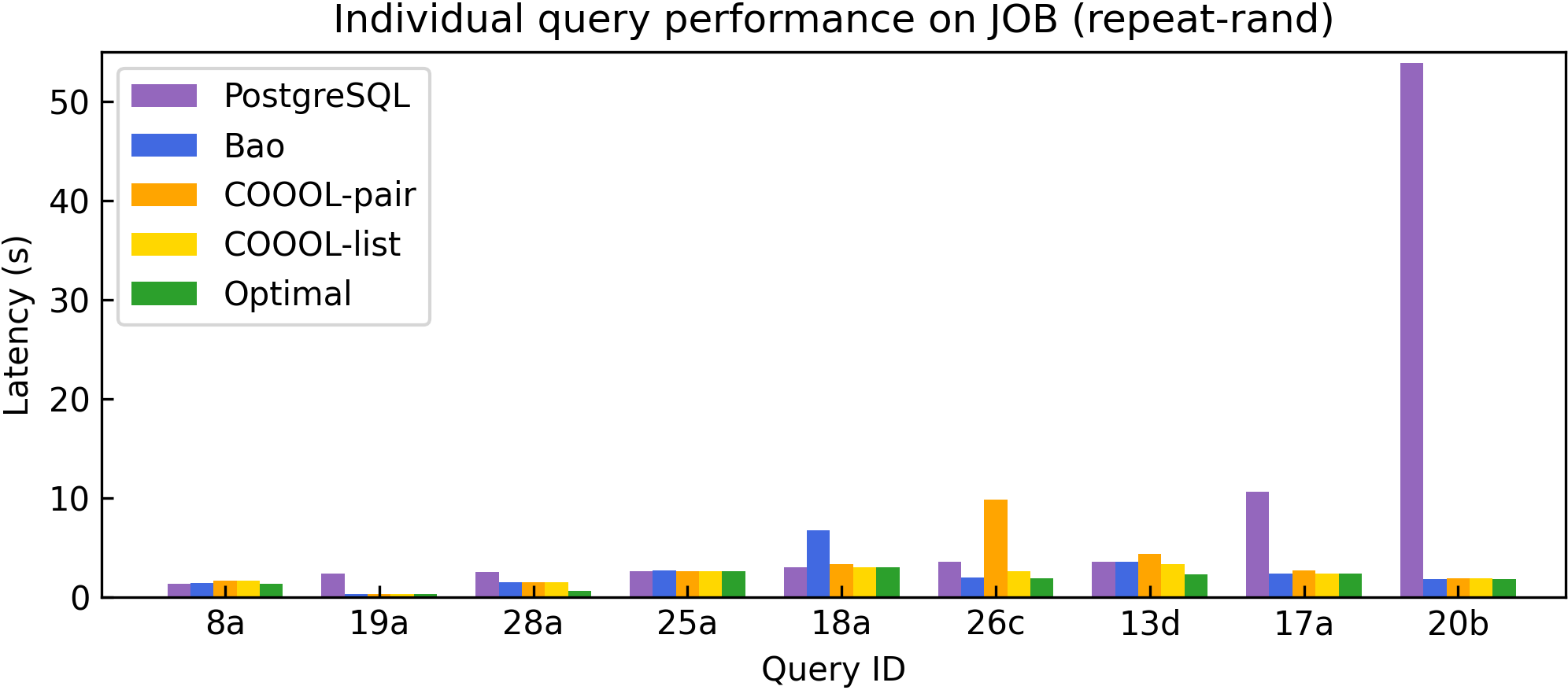}
         \caption{JOB repeat-rand}\label{fig:jobrepeatrand-unione}
     \end{subfigure}
     \hfill
     \begin{subfigure}[b]{0.45\textwidth}
         \centering
         \includegraphics[width=\textwidth]{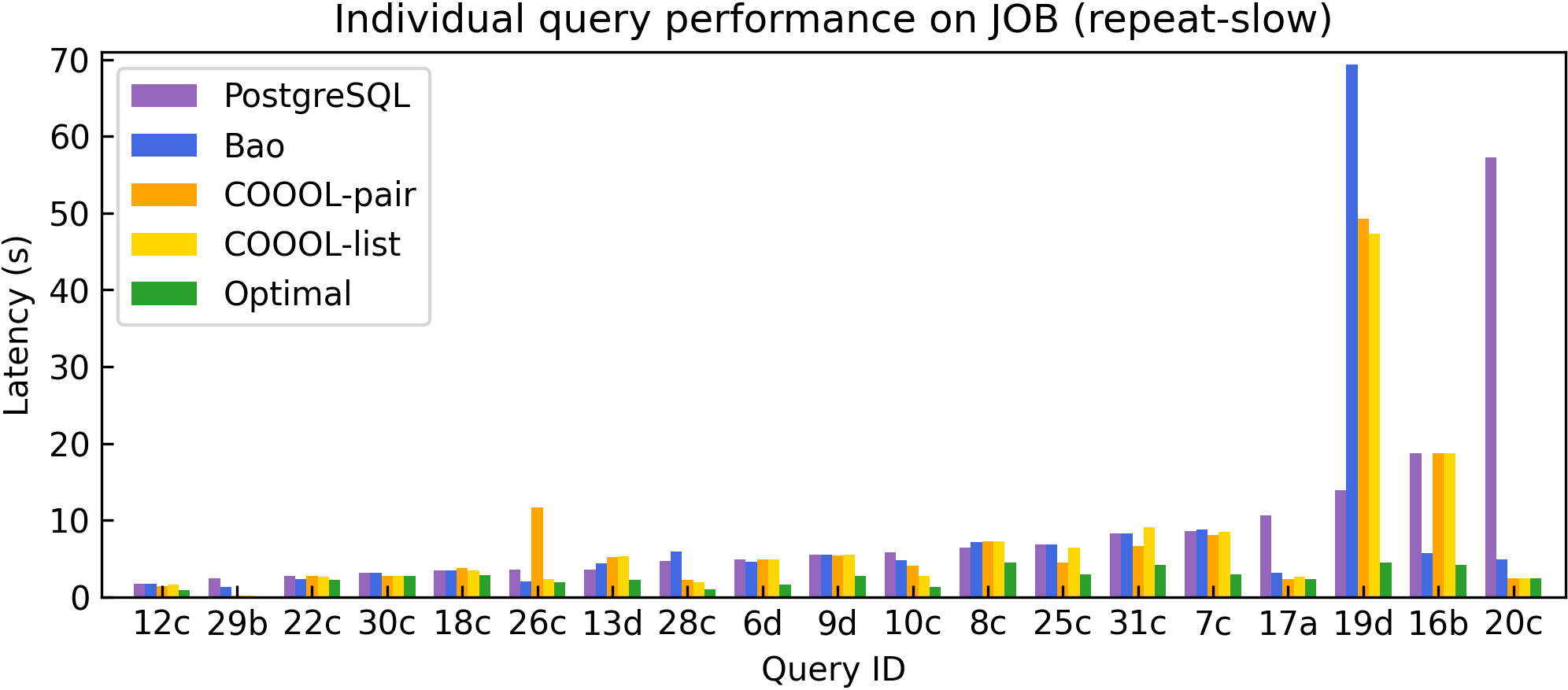}
         \caption{JOB repeat-slow}\label{fig:jobrepeatslow-unione}
     \end{subfigure}
     \hfill
     \begin{subfigure}[b]{0.45\textwidth}
         \centering
         \includegraphics[width=\textwidth]{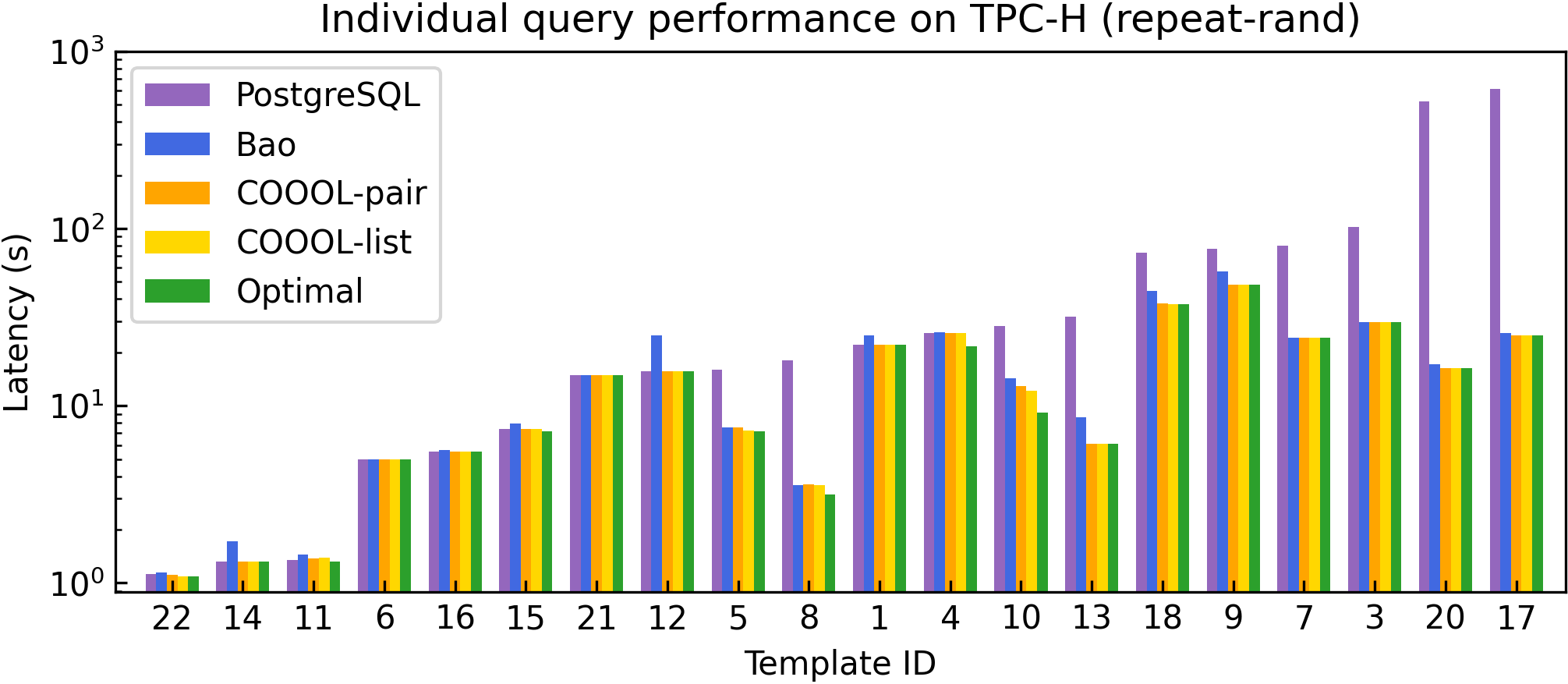}
         \caption{TPC-H repeat-rand}
     \end{subfigure}
     \hfill
     \begin{subfigure}[b]{0.45\textwidth}
         \centering
         \includegraphics[width=\textwidth]{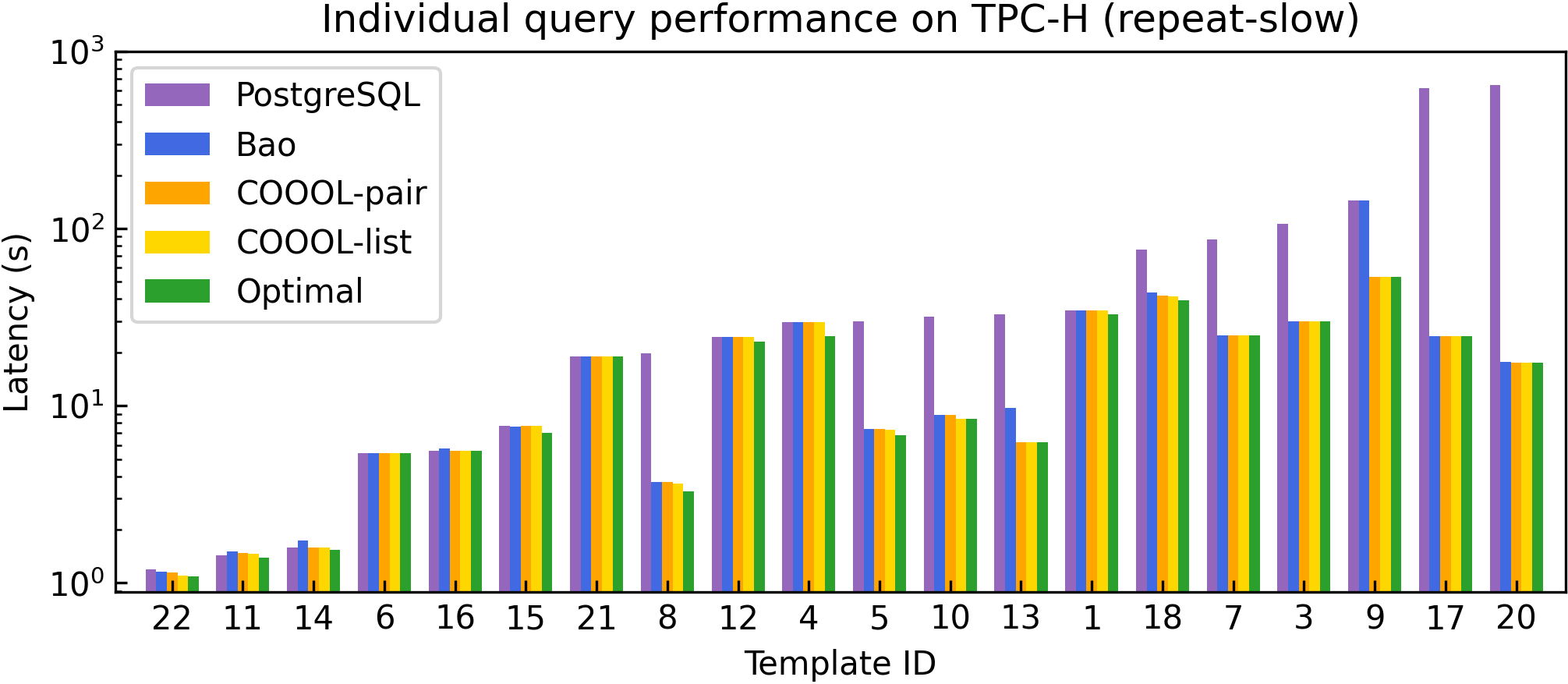}
         \caption{TPC-H repeat-slow}
     \end{subfigure}
     \hfill
    \caption{Individual query performance of the unified model.}
    \label{fig:unione_repeat}
\end{figure}

\subsection{Unified Model Performance (RQ3)}

\begin{table*}[]
    \centering
    \caption{Total query execution latency speedups of a unified model (trained on both JOB and TPC-H data) over PostgreSQL. The best performance on each workload is in boldface. ``$\uparrow$" indicates that the performance of the unified model is better than that trained on the corresponding single dataset.}
    \begin{tabular}{c|cccccccc}
    \toprule
        & \multicolumn{4}{c}{JOB} & \multicolumn{4}{c}{TPC-H} \\
        & adhoc-rand & adhoc-slow & repeat-rand & repeat-slow & adhoc-rand & adhoc-slow & repeat-rand & repeat-slow \\
    \midrule
        Bao & 0.80 & 0.92 & 2.79 & {1.21} & {5.77}$\uparrow$ & 1.93 & 4.83 & 4.31\\
        \textsf{COOOL}-list & 1.00 & 1.21 & \textbf{3.24}$\uparrow$ & \textbf{1.26} & {6.59}$\uparrow$ & {2.91} & \textbf{5.37} & \textbf{5.53}\\
        \textsf{COOOL}-pair & \textbf{1.06} & \textbf{1.71}$\uparrow$ & 2.90 & 1.21 & \textbf{6.73}$\uparrow$ & \textbf{3.92}$\uparrow$ & 5.34$\uparrow$ & {5.51}\\
        
    \bottomrule
    \end{tabular}
    \label{tab:one}
\end{table*}

This section focuses on the performance of models trained using both JOB and TPC-H datasets. Similar to the previous section, we consider four scenarios: ``adhoc-rand", ``adhoc-slow", ``repeat-rand", and ``repeat-slow" for each dataset. For each scenario, the training data of JOB and TPC-H were combined as the new training set and the model is respectively evaluated on JOB and TPC-H test sets. 

The overall results are summarized in Table~\ref{tab:one}. We observe that:
\begin{itemize}
    \item \textsf{COOOL}-pair performs the best in all ``adhoc" settings and \textsf{COOOL}-list performs the best in all ``repeat" settings. They have similar overall performances, and outperform Bao in almost all scenarios. 
    \item \textsf{COOOL}-pair has the most performance boost when the model is trained using both JOB and TPC-H datasets (unified model) compared with Table~\ref{tab:single}, especially on TPC-H where the unified model beats the single-instance model in three out of four scenarios. 
    \item Does a different training dataset help? We observe that JOB training data help improve the model performance on TPC-H test data, especially under ``adhoc" scenarios. While TPC-H training data do not help improve the model performance on JOB test data under most scenarios. We believe this is because JOB queries are more complicated (more nodes and larger depth than TPC-H queries as shown in Table~\ref{tab:plan-statistic}).
\end{itemize}

\begin{table}[t]
    \centering
    \caption{Number of regressions for Bao and \textsf{COOOL} compared with PostgreSQL (unified model)}
    \begin{tabular}{cccc}
    \toprule
        Setting & Bao & \textsf{COOOL}-list & \textsf{COOOL}-pair \\
    \midrule
        JOB repeat-rand & 23 & \textbf{12} & 17\\
        JOB repeat-slow & 20 & \textbf{9} & 11\\
        TPC-H repeat-rand & 8 & \textbf{1} & \textbf{1}\\
        TPC-H repeat-slow & 4 & \textbf{1} & \textbf{1}\\
    \bottomrule
    \end{tabular}
    \label{tab:unified-regression}
\end{table}

The individual query performance in ``repeat" settings of the unified models are shown in Figure \ref{fig:unione_repeat}, where we also depict queries with an execution latency greater than 1s on PostgreSQL. Combined with the observation in Figure \ref{fig:unione_repeat}, we can get the following conclusions:
\begin{itemize}
    \item The performances of both \textsf{COOOL} approaches are close to optimal. This echoes our observation that JOB training data help improve the model performances on TPC-H. 
    \item On the other hand, the model performances on JOB test data are hurt by the TPC-H training data. For example, both Bao and \textsf{COOOL} methods have poor performances on query 19d in Figure \ref{fig:jobrepeatslow-unione}. 
    \item In terms of number of regressions in excution time, we list the results in Table~\ref{tab:unified-regression}. Both \textsf{COOOL}-pair and \textsf{COOOL}-list perform better than Bao, and \textsf{COOOL}-list has the lowest number of query regressions for these settings.
\end{itemize}

\paragraph{Summary on Model Performances} Different data distributions in different datasets is a challenge for a unified query optimization model, but \textsf{COOOL} models are able to alleviate this issue to learn a unified model better than the state-of-the-art regression approach to speed up total query execution, alleviate individual query regression, and optimize slow queries. Our experiments have shown the overwhelming advantages of our proposed models. For the single instance scenarios, \textsf{COOOL}-list achieves the best performance in total query execution speedups while \textsf{COOOL}-pair is the best to reduce the number of individual query regressions. When a unified model is trained using multiple datasets, \textsf{COOOL}-list performs the best in ``repeat" settings, while \textsf{COOOL}-pair is the best in ``adhoc" settings and has no total query execution performance regression. \textsf{COOOL}-pair is the best approach among the three models in terms of stability. 

\begin{table}[t]
    \centering
    \caption{Comparison of training time required for convergence in the ``adhoc-slow" setting}
    \begin{tabular}{cccccccccc}
    \toprule
         & JOB & TPC-H & Unified &\\
    \midrule
        Bao &  119.5s & 34.5s & 265.9s\\
        \textsf{COOOL}-list &167.0s & 159.8s & 317.1s \\
        \textsf{COOOL}-pair &493.6s & 222.8s & 894.6s \\
    \bottomrule
    \end{tabular}
    \label{tab:time}
\end{table}

\begin{figure*}[ht]
     \centering
     \begin{subfigure}[b]{0.3\textwidth}
         \centering
         \includegraphics[width=\textwidth]{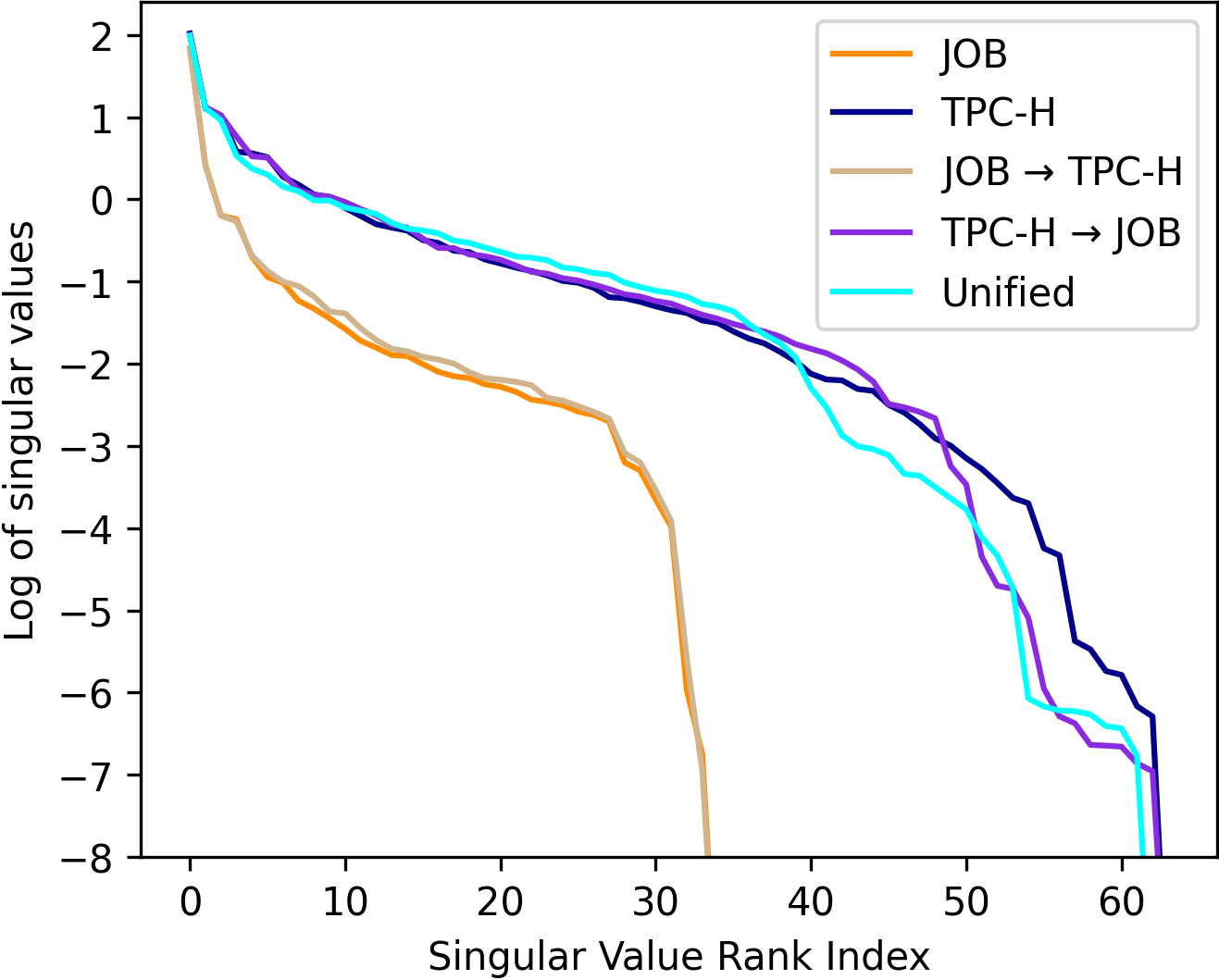}
         \caption{Bao}
         \label{fig:bao-svd}
     \end{subfigure}
     \hfill
     \begin{subfigure}[b]{0.3\textwidth}
         \centering
         \includegraphics[width=\textwidth]{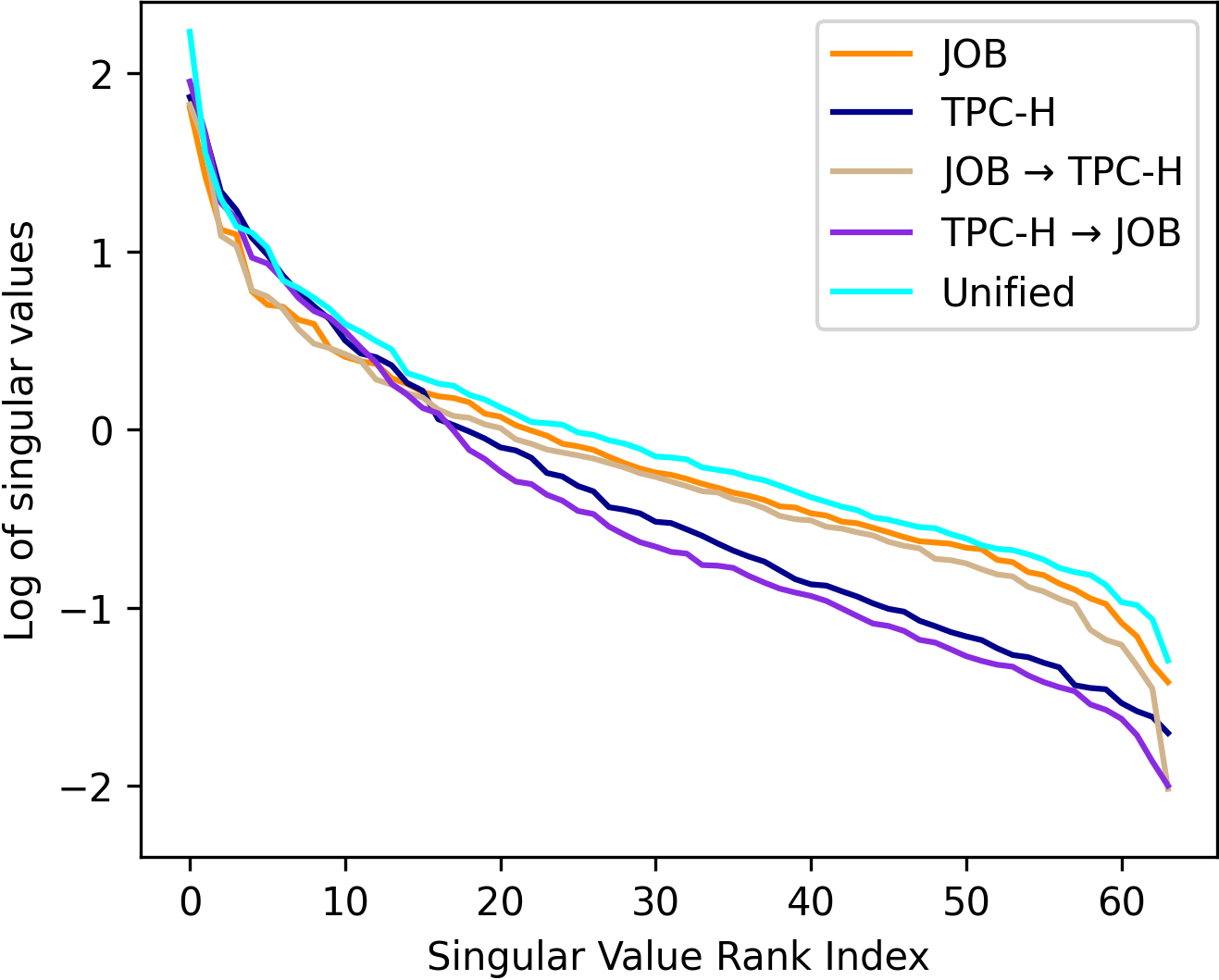}
         \caption{\textsf{COOOL}-pair}
         \label{fig:coool-pair-svd}
     \end{subfigure}
     \hfill
     \begin{subfigure}[b]{0.3\textwidth}
         \centering
         \includegraphics[width=\textwidth]{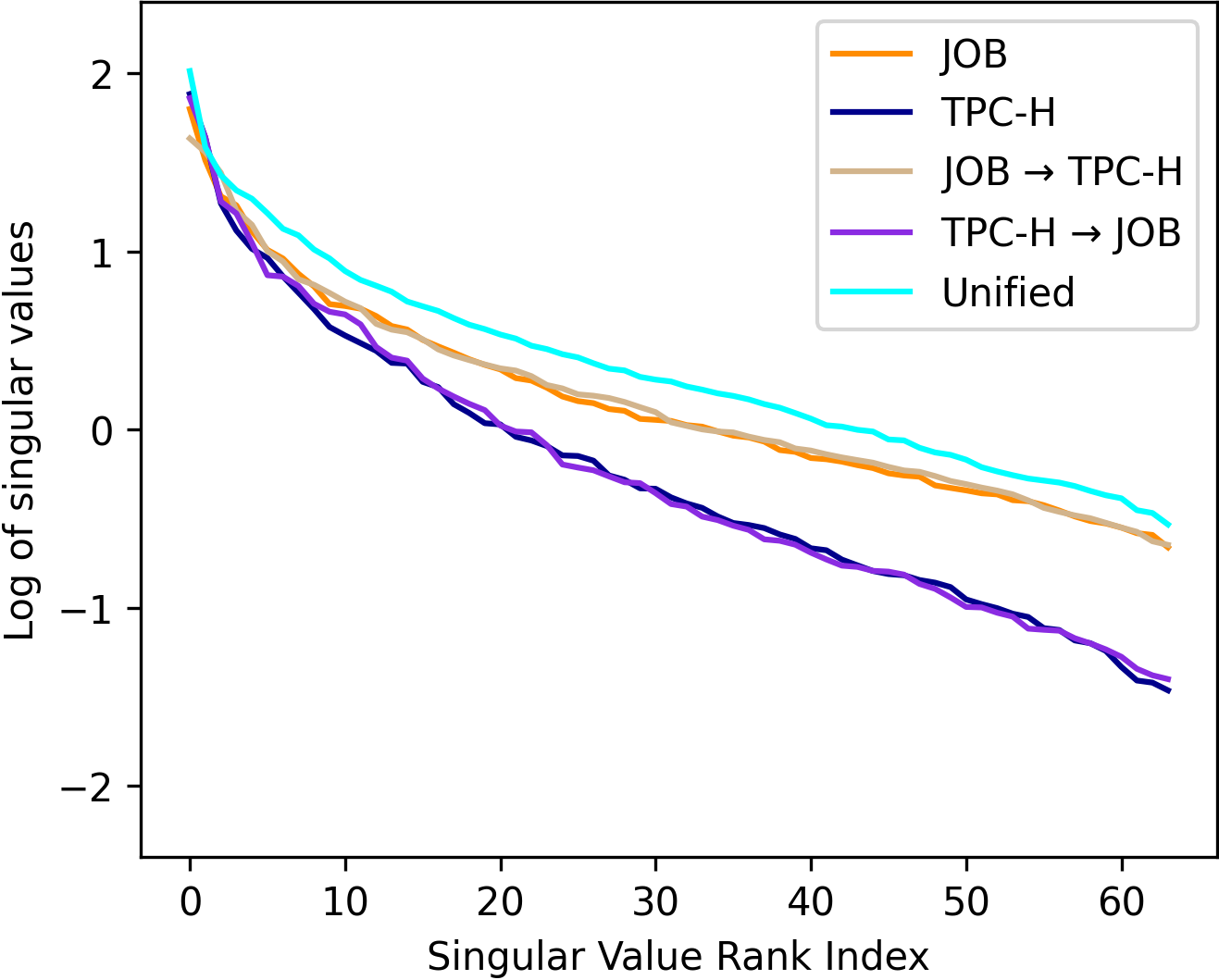}
         \caption{\textsf{COOOL}-list}
         \label{fig:coool-list-svd}
     \end{subfigure}
     \hfill
    \caption{Singular value spectrum of the plan embedding space for Bao and \textsf{COOOL} in ``adhoc-slow" setting of the two workloads in the scenarios of the single instance, workload transfer, and unified model.}
    \label{fig:singular}
\end{figure*}

\subsection{Model Comparison and Analysis (RQ4)}

We have conducted extensive experiments and obtained promising results. But it is not clear how the ranking strategies affect the model training. This section aims to provide insight on why the two \textsf{COOOL} models outperforms Bao, especially for the unified model training. 

In \textsf{COOOL} methods, we use exactly the same plan representation model as Bao to confirm that the improvements are brought by our LTR methods. We analyze the ``adhoc-slow" setting of the two workloads in this section.

\subsubsection{Model efficiency.} 
We compare the space and time efficiency of our approaches and state-of-the-art method.
\begin{itemize}
    \item {{Space complexity.}} The two \textsf{COOOL} models use tree convolutional neural networks with the same structure and hidden sizes as Bao, the number of parameters for all of them is 132,353, i.e., 529,412 bytes. The storage consumption of the model is about 0.5MB, which is efficient enough in practice.
    \item {{The number of training samples.}} Since there are three different strategies involved in this paper, it is difficult to directly derive their time complexity. So we first show the number of training samples.
    Let $N$ denote the number of training queries. And there are $n$ candidate hint sets in $\mathcal{H}$ for each query. The time complexity of Bao is $Nn$. Let $m_i$ denote the number of unique query plans for each query given $\mathcal{H}$, where $i\in\{1, 2, \ldots, N\}$ and $m_i\leq n$. Because we use a full rank-breaking strategy in \textsf{COOOL}-pair, the number of sample is $\Theta(\sum_{i=1}^N\frac{m_i(m_i-1)}{2})=O(Nn^2)$. Since $n$ is a given constant in practice, the complexity is acceptable. For \textsf{COOOL}-list, there are $N$ training samples.
    \item {{Training time consumption.}} To intuitively demonstrate the time efficiency of the three methods, we summarize the average training time required for convergence in the ``adhoc-slow" settings, as shown in Table \ref{tab:time}. It is obvious that \textsf{COOOL} indeed needs more training time for convergence than Bao. Besides, \textsf{COOOL}-pair requires much more time for convergence than \textsf{COOOL}-list.
    \item {Conclusion.} Notwithstanding \textsf{COOOL}-pair and \textsf{COOOL}-list require longer convergence time than Bao, they have the same model inference efficiency and the number parameters as Bao. Therefore, we can conclude that \textsf{COOOL} is a practical method.
\end{itemize}

\subsubsection{A representation learning perspective for plan tree embeddings.} \label{sec:RepresentationLearning}
Regression-based methods usually limit the model output domain in an interval depending on the normalization methods. By contrast, ranking-based methods do not limit the absolute value of model outputs, which may help the model learn better plan representations. This section aims to provide some insight on why \textsf{COOOL} outperforms Bao from a representation learning perspective. 

Let an $h$-dimensional vector $\mathbf{z_i}$ denote the embedding for query plan $i$. We compute the covariance matrix $C\in\mathbb{R}^{h\times h}$ for all plan embeddings obtained by the model. Formally, 
$$C=\frac{1}{M}\sum_{i=1}^M(\mathbf{z_i}-\overline{\mathbf{z}})(\mathbf{z_i}-\overline{\mathbf{z}})^\top,$$
where $M$ is the number of plans, $\overline{\mathbf{z}}=\frac{1}{M}\sum_{i=1}^M\mathbf{z_i}$. Then, we apply singular value decomposition on $C$ s.t. $C=USV^\top$, where $S=diag(\sigma^k)$. Then we can obtain the singular value spectrum in sorted order and logarithmic scale ($\lg(\sigma^k)$). We depict the singular value spectrum of the plan embedding space of the three models learned in different scenarios in ``adhoc-slow" setting, as shown in Figure \ref{fig:singular}. We have the following observations.

\begin{itemize}
    \item In Figure \ref{fig:bao-svd}, we observe a significant drop in singular values on a logarithmic scale ($\lg(\sigma^k)$) in all tasks. The curve approaches zero (less than 1e-7) in the spectrum indicating that a dimensional collapse \cite{dimensionalCollapse} occurs in Bao's embedding space in each of Bao's experimental scenarios, and the model learned on JOB even collapsed in half of the plan embedding space. The dimensional collapses result in Bao's plan embedding vectors only spanning a lower-dimensional subspace, which may harm the representation ability of the model and therefore hurt the performance. Moreover, the models in the single instance scenario on different datasets result in a different number of collapsed dimensions. Based on the workload transfer scenario curves, we observe that the plan embedding space of the target workload spans the same number of dimensions as the source workload. So we can conclude that the embedding space is determined by the training data.
    \item In Figures \ref{fig:coool-pair-svd} and \ref{fig:coool-list-svd}, singular values of the \textsf{COOOL} approaches does not drop abruptly, meaning that there is no collapse in their plan embedding space. The plan embedding methods of Bao and COOOL are the same, and the comparison reflects that the embedding spaces learned from the regression framework and ranking strategies are significantly different.
    \item The spectrum of Bao's unified model is closer to the spectrum of the learned representation from TPC-H data than that learned from JOB data, whereas the spectra of \textsf{COOOL}-pair and \textsf{COOOL}-list are closer to the spectrum of the representation from JOB data than that learned from TPC-H data. In fact, the number of training samples of TPC-H is greater than that of JOB, and JOB is more complicated than TPC-H (see Table~\ref{tab:plan-statistic}). When learning from multiple datasets, Bao's performance is affected more by the datasets with more samples. By contrast, the two \textsf{COOOL} methods are able to learn from sophisticated query plans.
    \item Singular values are related to the latent information in the covariance matrix. We can get three conclusions from this perspective. First, the overall spectrums of Bao in different settings are lower than \textsf{COOOL} in terms of absolute singular value, which indicates the representation ability of Bao is worse than \textsf{COOOL}. Second, the curve of Bao's unified model is not the highest in most dimensions, while that of the two \textsf{COOOL} models are just the opposite. It may reveal that the ranking strategies can help models learn latent patterns from two workloads but the regression framework cannot. Third, the range and the trend of the curves of \textsf{COOOL}-pair and \textsf{COOOL}-list are different, which may result in the different performances of the two models in different scenarios.
    \item The dimensional collapse issue may not greatly hurt single instance optimization since the embedding space is learned from one workload and the latent collapsed dimensions are fixed when the model converges. Because the collapsed dimensions may be different in different datasets, there may be orthogonal situations, which may cause the subspace representation learned in one dataset to be noise in another, resulting in unstable performance. Therefore, it may limit the scalability of machine learning query optimization models.
\end{itemize}

To sum up, dimensional collapse in plan tree embedding space is a challenge for machine learning query optimization models in maintaining a unified model to learn from different workloads. The two variants of \textsf{COOOL} have the exact same model as Bao but consistently outperform Bao in almost all scenarios and settings. We provide some insight on why \textsf{COOOL} methods outperform Bao from the query plan representation perspective. It may guide researchers to further improve the performance of ML models for query optimization. %

\section{Related work}
\label{RelatedWork}
\subsection{Machine Learning for Query Optimization}
Traditional cost-based query optimizers aim to select the candidate plan with the minimum estimated cost, where cost represents the execution latency or other user-defined resource consumption metrics. Various techniques have been proposed in cost-based optimizers \cite{cbohistogram_ioannidis1995balancing,cbosketch_cormode2011synopses,cboprobabilistic_cormode2010histograms}, such as sketches, histograms, probability models, etc. Besides, they work with different assumptions (e.g., attribute value independence \cite{cboindependenceassum_christodoulakis1984implications}, uniformity \cite{cbouniformassum_ioannidis1993optimal}, data independence \cite{cboindependenceassum2_deshpande2001independence}) and when these cannot be met, the techniques usually fall back to an educated guess \cite{cbowoassumption_poosala1997selectivity}. Traditional optimizers have been studied for decades \cite{oldwork_selinger1979access}, focusing on manually-crafted and heuristic methods to predict costs, estimate cardinality, and generate plans \cite{cbosurvey_ioannidis1996query}. However, the effect of heuristics depends on data distribution, cardinality estimation, and cost modeling have been proven difficult to solve \cite{provequeryhard_lohman2014query}. Consequently, cost-based optimizers may generate poor performance plans.

Therefore, the database community has attempted to apply Machine Learning (ML) techniques to solve these issues. For example, some efforts introduced Reinforcement Learning (RL) and Monte Carlo Tree Search (MCTS) to optimize the join order selection task \cite{dq_krishnan2018learning,rejoin_marcus2018deep,trummer2021skinnerdb,alphajoin_zhang2020alphajoin}, and some studies use neural networks to accomplish cost modeling and cardinality estimation \cite{liu2015cardinality,e2ecost_19_sunji,naru_yang2019deep,yang2020neurocard}. These works depict that the elaborated ML models can improve the performance of parts of those components in query optimization. However, they may not improve the performance of optimizers since none of them demonstrate that the performance improvement of a single component can actually bring a better query plan~\cite{cardinalityandquery_leis2018query}.

In recent years, some studies have attempted to build end-to-end ML models for query optimization \cite{neo,marcus2022bao,balsa_SIGMOD22}. They all use a TCNN to predict the cost/latency of query plans and leverage a deep reinforcement learning framework to train the model. 
Though they have made improvements compared with the previous works of replacing some components of the optimizer with ML models, the evaluation of an optimizer should be measured from multiple dimensions \cite{aidbsurvey_tsesmelis2022database}, and some of the most concerned parts are \textit{practical}, \textit{data/schema agnostic}, \textit{explainability}, \textit{performance}, \textit{scalability}, etc. Neo \cite{neo} learns row vector embeddings from tables, so it needs to maintain row vector embeddings for different data. Balsa \cite{balsa_SIGMOD22} can not treat advanced SQL features (e.g., sub-queries) due to its dependency, which limits its scalability. Furthermore, Neo and Balsa are less practical than Bao \cite{marcus2022bao}. Bao introduced a novel approach to end-to-end query optimization that recommends per-query hints over an existing optimizer. SQL hints can limit the search space of existing optimizers and each set of hints corresponds to a plan, which makes it practical in real scenarios.

\subsection{Learning-To-Rank}
Learning-To-Rank (LTR) is an active research topic \cite{yu2023depth} and has many applications in information retrieval~\cite{Liu09:Learning}, meta-search engines~\cite{Dwork01:Rank}, recommendation systems~\cite{Karatzoglou13:Learning}, preference learning~\cite{Zhao22:Learning}, etc. A search engine or a recommendation system usually need to rank a large and variable number of items (webpages, movies, for example). The SQL hint recommendation problem is more like the preference learning problem, where a relatively small and fixed number of items (SQL hints in our case), are to be ranked for each user (SQL query in our case). Plackett-Luce model~\cite{Plackett75:Analysis,Luce59:Individual}, which was later used as a listwise loss function in information retrieval~\cite{Liu09:Learning} and softmax function in classification tasks~\cite{Gao17:Properties}, is one of the most popular models in preference learning. It is a listwise model, but can be learned efficiently with pairwise methods~\cite{Khetan16:Data,Zhao18:Composite}, by breaking the full rankings into pairwise comparisons and minimizing a pairwise loss function. Similar to \cite{Khetan16:Data} and \cite{Zhao18:Composite}, we maximize the marginal pairwise likelihood for its similicity. 

\FloatBarrier
\section{Conclusion}
\label{Conclusion}

In this paper, we propose \textsf{COOOL} that predicts cost orders of query plans to cooperate with DBMS by LTR techniques to recommend SQL hints for query optimization. \textsf{COOOL} has pairwise and listwise methods, each of which can improve overall query execution speed especially on slow queries and alleviate individual query regression. Moreover, we can maintain a unified model to improve query plan selections by the proposed method, and we shed some light on why both \textsf{COOOL}-list and \textsf{COOOL}-pair approaches outperform Bao from the representation learning perspective. 

\textsf{COOOL} makes a step forward to maintaining a unified end-to-end query optimization ML model on two datasets, and the elaborated analysis may provide preliminary for large-scale pre-trained query optimization models. While \textsf{COOOL} methods are promising, they cannot estimate the cost of the recommended query plan, or quantitatively compare the costs of two query plans.

For future work, we plan to investigate the evaluation metrics for ranking candidate plans that differ by multiple orders of magnitude in execution latency for query optimization, which facilitates introducing state-of-the-art LTR techniques. LTR is still an active research field in recent years, so there are considerable future works to be explored in query optimization. Besides, developing large-scale pre-trained query optimization models and quantitatively compare the costs of different query plans accurately are also challenging future directions.

\balance

\bibliographystyle{ACM-Reference-Format}
\bibliography{ref,ltr,ai4db}


\begin{thebibliography}{47}


\ifx \showCODEN    \undefined \def \showCODEN     #1{\unskip}     \fi
\ifx \showDOI      \undefined \def \showDOI       #1{#1}\fi
\ifx \showISBNx    \undefined \def \showISBNx     #1{\unskip}     \fi
\ifx \showISBNxiii \undefined \def \showISBNxiii  #1{\unskip}     \fi
\ifx \showISSN     \undefined \def \showISSN      #1{\unskip}     \fi
\ifx \showLCCN     \undefined \def \showLCCN      #1{\unskip}     \fi
\ifx \shownote     \undefined \def \shownote      #1{#1}          \fi
\ifx \showarticletitle \undefined \def \showarticletitle #1{#1}   \fi
\ifx \showURL      \undefined \def \showURL       {\relax}        \fi
\providecommand\bibfield[2]{#2}
\providecommand\bibinfo[2]{#2}
\providecommand\natexlab[1]{#1}
\providecommand\showeprint[2][]{arXiv:#2}

\bibitem[\protect\citeauthoryear{Akdere, {\c{C}}etintemel, Riondato, Upfal, and
  Zdonik}{Akdere et~al\mbox{.}}{2012}]%
        {akdere2012learning}
\bibfield{author}{\bibinfo{person}{Mert Akdere}, \bibinfo{person}{Ugur
  {\c{C}}etintemel}, \bibinfo{person}{Matteo Riondato}, \bibinfo{person}{Eli
  Upfal}, {and} \bibinfo{person}{Stanley~B Zdonik}.}
  \bibinfo{year}{2012}\natexlab{}.
\newblock \showarticletitle{Learning-based query performance modeling and
  prediction}. In \bibinfo{booktitle}{\emph{2012 IEEE 28th International
  Conference on Data Engineering}}. IEEE, \bibinfo{pages}{390--401}.
\newblock


\bibitem[\protect\citeauthoryear{Azari~Soufiani, Chen, Parkes, and
  Xia}{Azari~Soufiani et~al\mbox{.}}{2013}]%
        {Azari13:Generalized}
\bibfield{author}{\bibinfo{person}{Hossein Azari~Soufiani},
  \bibinfo{person}{William Chen}, \bibinfo{person}{David~C Parkes}, {and}
  \bibinfo{person}{Lirong Xia}.} \bibinfo{year}{2013}\natexlab{}.
\newblock \showarticletitle{Generalized method-of-moments for rank
  aggregation}.
\newblock \bibinfo{journal}{\emph{Advances in Neural Information Processing
  Systems}}  \bibinfo{volume}{26} (\bibinfo{year}{2013}).
\newblock


\bibitem[\protect\citeauthoryear{Christodoulakis}{Christodoulakis}{1984}]%
        {cboindependenceassum_christodoulakis1984implications}
\bibfield{author}{\bibinfo{person}{Stavros Christodoulakis}.}
  \bibinfo{year}{1984}\natexlab{}.
\newblock \showarticletitle{Implications of certain assumptions in database
  performance evauation}.
\newblock \bibinfo{journal}{\emph{ACM Transactions on Database Systems (TODS)}}
  \bibinfo{volume}{9}, \bibinfo{number}{2} (\bibinfo{year}{1984}),
  \bibinfo{pages}{163--186}.
\newblock


\bibitem[\protect\citeauthoryear{Cormode and Garofalakis}{Cormode and
  Garofalakis}{2010}]%
        {cboprobabilistic_cormode2010histograms}
\bibfield{author}{\bibinfo{person}{Graham Cormode} {and} \bibinfo{person}{Minos
  Garofalakis}.} \bibinfo{year}{2010}\natexlab{}.
\newblock \showarticletitle{Histograms and wavelets on probabilistic data}.
\newblock \bibinfo{journal}{\emph{IEEE Transactions on Knowledge and Data
  Engineering}} \bibinfo{volume}{22}, \bibinfo{number}{8}
  (\bibinfo{year}{2010}), \bibinfo{pages}{1142--1157}.
\newblock


\bibitem[\protect\citeauthoryear{Cormode, Garofalakis, Haas, Jermaine,
  et~al\mbox{.}}{Cormode et~al\mbox{.}}{2011}]%
        {cbosketch_cormode2011synopses}
\bibfield{author}{\bibinfo{person}{Graham Cormode}, \bibinfo{person}{Minos
  Garofalakis}, \bibinfo{person}{Peter~J Haas}, \bibinfo{person}{Chris
  Jermaine}, {et~al\mbox{.}}} \bibinfo{year}{2011}\natexlab{}.
\newblock \showarticletitle{Synopses for massive data: Samples, histograms,
  wavelets, sketches}.
\newblock \bibinfo{journal}{\emph{Foundations and Trends{\textregistered} in
  Databases}} \bibinfo{volume}{4}, \bibinfo{number}{1--3}
  (\bibinfo{year}{2011}), \bibinfo{pages}{1--294}.
\newblock


\bibitem[\protect\citeauthoryear{Deshpande, Garofalakis, and Rastogi}{Deshpande
  et~al\mbox{.}}{2001}]%
        {cboindependenceassum2_deshpande2001independence}
\bibfield{author}{\bibinfo{person}{Amol Deshpande}, \bibinfo{person}{Minos
  Garofalakis}, {and} \bibinfo{person}{Rajeev Rastogi}.}
  \bibinfo{year}{2001}\natexlab{}.
\newblock \showarticletitle{Independence is good: Dependency-based histogram
  synopses for high-dimensional data}.
\newblock \bibinfo{journal}{\emph{ACM SIGMOD Record}} \bibinfo{volume}{30},
  \bibinfo{number}{2} (\bibinfo{year}{2001}), \bibinfo{pages}{199--210}.
\newblock


\bibitem[\protect\citeauthoryear{Dwork, Kumar, Naor, and Sivakumar}{Dwork
  et~al\mbox{.}}{2001}]%
        {Dwork01:Rank}
\bibfield{author}{\bibinfo{person}{Cynthia Dwork}, \bibinfo{person}{Ravi
  Kumar}, \bibinfo{person}{Moni Naor}, {and} \bibinfo{person}{Dandapani
  Sivakumar}.} \bibinfo{year}{2001}\natexlab{}.
\newblock \showarticletitle{Rank aggregation methods for the web}. In
  \bibinfo{booktitle}{\emph{Proceedings of the 10th international conference on
  World Wide Web}}. \bibinfo{pages}{613--622}.
\newblock


\bibitem[\protect\citeauthoryear{Gao and Pavel}{Gao and Pavel}{2017}]%
        {Gao17:Properties}
\bibfield{author}{\bibinfo{person}{Bolin Gao} {and} \bibinfo{person}{Lacra
  Pavel}.} \bibinfo{year}{2017}\natexlab{}.
\newblock \showarticletitle{On the properties of the softmax function with
  application in game theory and reinforcement learning}.
\newblock \bibinfo{journal}{\emph{arXiv preprint arXiv:1704.00805}}
  (\bibinfo{year}{2017}).
\newblock


\bibitem[\protect\citeauthoryear{Hua, Wang, Xue, Ren, Wang, and Zhao}{Hua
  et~al\mbox{.}}{2021}]%
        {dimensionalCollapse}
\bibfield{author}{\bibinfo{person}{Tianyu Hua}, \bibinfo{person}{Wenxiao Wang},
  \bibinfo{person}{Zihui Xue}, \bibinfo{person}{Sucheng Ren},
  \bibinfo{person}{Yue Wang}, {and} \bibinfo{person}{Hang Zhao}.}
  \bibinfo{year}{2021}\natexlab{}.
\newblock \showarticletitle{On Feature Decorrelation in Self-Supervised
  Learning}. In \bibinfo{booktitle}{\emph{2021 {IEEE/CVF} International
  Conference on Computer Vision, {ICCV} 2021, Montreal, QC, Canada, October
  10-17, 2021}}. \bibinfo{publisher}{{IEEE}}, \bibinfo{pages}{9578--9588}.
\newblock
\urldef\tempurl%
\url{https://doi.org/10.1109/ICCV48922.2021.00946}
\showDOI{\tempurl}


\bibitem[\protect\citeauthoryear{Ioannidis}{Ioannidis}{1996}]%
        {cbosurvey_ioannidis1996query}
\bibfield{author}{\bibinfo{person}{Yannis~E Ioannidis}.}
  \bibinfo{year}{1996}\natexlab{}.
\newblock \showarticletitle{Query optimization}.
\newblock \bibinfo{journal}{\emph{ACM Computing Surveys (CSUR)}}
  \bibinfo{volume}{28}, \bibinfo{number}{1} (\bibinfo{year}{1996}),
  \bibinfo{pages}{121--123}.
\newblock


\bibitem[\protect\citeauthoryear{Ioannidis and Christodoulakis}{Ioannidis and
  Christodoulakis}{1993}]%
        {cbouniformassum_ioannidis1993optimal}
\bibfield{author}{\bibinfo{person}{Yannis~E Ioannidis} {and}
  \bibinfo{person}{Stavros Christodoulakis}.} \bibinfo{year}{1993}\natexlab{}.
\newblock \showarticletitle{Optimal histograms for limiting worst-case error
  propagation in the size of join results}.
\newblock \bibinfo{journal}{\emph{ACM Transactions on Database Systems (TODS)}}
  \bibinfo{volume}{18}, \bibinfo{number}{4} (\bibinfo{year}{1993}),
  \bibinfo{pages}{709--748}.
\newblock


\bibitem[\protect\citeauthoryear{Ioannidis and Poosala}{Ioannidis and
  Poosala}{1995}]%
        {cbohistogram_ioannidis1995balancing}
\bibfield{author}{\bibinfo{person}{Yannis~E Ioannidis} {and}
  \bibinfo{person}{Viswanath Poosala}.} \bibinfo{year}{1995}\natexlab{}.
\newblock \showarticletitle{Balancing histogram optimality and practicality for
  query result size estimation}.
\newblock \bibinfo{journal}{\emph{Acm Sigmod Record}} \bibinfo{volume}{24},
  \bibinfo{number}{2} (\bibinfo{year}{1995}), \bibinfo{pages}{233--244}.
\newblock


\bibitem[\protect\citeauthoryear{Karatzoglou, Baltrunas, and Shi}{Karatzoglou
  et~al\mbox{.}}{2013}]%
        {Karatzoglou13:Learning}
\bibfield{author}{\bibinfo{person}{Alexandros Karatzoglou},
  \bibinfo{person}{Linas Baltrunas}, {and} \bibinfo{person}{Yue Shi}.}
  \bibinfo{year}{2013}\natexlab{}.
\newblock \showarticletitle{Learning to rank for recommender systems}. In
  \bibinfo{booktitle}{\emph{Proceedings of the 7th ACM Conference on
  Recommender Systems}}. \bibinfo{pages}{493--494}.
\newblock


\bibitem[\protect\citeauthoryear{Khetan and Oh}{Khetan and Oh}{2016}]%
        {Khetan16:Data}
\bibfield{author}{\bibinfo{person}{Ashish Khetan} {and}
  \bibinfo{person}{Sewoong Oh}.} \bibinfo{year}{2016}\natexlab{}.
\newblock \showarticletitle{Data-driven rank breaking for efficient rank
  aggregation}. In \bibinfo{booktitle}{\emph{International Conference on
  Machine Learning}}. PMLR, \bibinfo{pages}{89--98}.
\newblock


\bibitem[\protect\citeauthoryear{Kingma and Ba}{Kingma and Ba}{2015}]%
        {adam_kingma2014adam}
\bibfield{author}{\bibinfo{person}{Diederik~P. Kingma} {and}
  \bibinfo{person}{Jimmy Ba}.} \bibinfo{year}{2015}\natexlab{}.
\newblock \showarticletitle{Adam: {A} Method for Stochastic Optimization}. In
  \bibinfo{booktitle}{\emph{3rd International Conference on Learning
  Representations, {ICLR} 2015, San Diego, CA, USA, May 7-9, 2015, Conference
  Track Proceedings}}.
\newblock


\bibitem[\protect\citeauthoryear{Krishnan, Yang, Goldberg, Hellerstein, and
  Stoica}{Krishnan et~al\mbox{.}}{2018}]%
        {dq_krishnan2018learning}
\bibfield{author}{\bibinfo{person}{Sanjay Krishnan}, \bibinfo{person}{Zongheng
  Yang}, \bibinfo{person}{Ken Goldberg}, \bibinfo{person}{Joseph Hellerstein},
  {and} \bibinfo{person}{Ion Stoica}.} \bibinfo{year}{2018}\natexlab{}.
\newblock \showarticletitle{Learning to optimize join queries with deep
  reinforcement learning}.
\newblock \bibinfo{journal}{\emph{arXiv preprint arXiv:1808.03196}}
  (\bibinfo{year}{2018}).
\newblock


\bibitem[\protect\citeauthoryear{Leis, Gubichev, Mirchev, Boncz, Kemper, and
  Neumann}{Leis et~al\mbox{.}}{2015}]%
        {JOB_leis2015good}
\bibfield{author}{\bibinfo{person}{Viktor Leis}, \bibinfo{person}{Andrey
  Gubichev}, \bibinfo{person}{Atanas Mirchev}, \bibinfo{person}{Peter Boncz},
  \bibinfo{person}{Alfons Kemper}, {and} \bibinfo{person}{Thomas Neumann}.}
  \bibinfo{year}{2015}\natexlab{}.
\newblock \showarticletitle{How good are query optimizers, really?}
\newblock \bibinfo{journal}{\emph{Proceedings of the VLDB Endowment}}
  \bibinfo{volume}{9}, \bibinfo{number}{3} (\bibinfo{year}{2015}),
  \bibinfo{pages}{204--215}.
\newblock


\bibitem[\protect\citeauthoryear{Leis, Radke, Gubichev, Mirchev, Boncz, Kemper,
  and Neumann}{Leis et~al\mbox{.}}{2018}]%
        {cardinalityandquery_leis2018query}
\bibfield{author}{\bibinfo{person}{Viktor Leis}, \bibinfo{person}{Bernhard
  Radke}, \bibinfo{person}{Andrey Gubichev}, \bibinfo{person}{Atanas Mirchev},
  \bibinfo{person}{Peter Boncz}, \bibinfo{person}{Alfons Kemper}, {and}
  \bibinfo{person}{Thomas Neumann}.} \bibinfo{year}{2018}\natexlab{}.
\newblock \showarticletitle{Query optimization through the looking glass, and
  what we found running the join order benchmark}.
\newblock \bibinfo{journal}{\emph{The VLDB Journal}}  \bibinfo{volume}{27}
  (\bibinfo{year}{2018}), \bibinfo{pages}{643--668}.
\newblock


\bibitem[\protect\citeauthoryear{Liu, Xu, Yu, Corvinelli, and Zuzarte}{Liu
  et~al\mbox{.}}{2015}]%
        {liu2015cardinality}
\bibfield{author}{\bibinfo{person}{Henry Liu}, \bibinfo{person}{Mingbin Xu},
  \bibinfo{person}{Ziting Yu}, \bibinfo{person}{Vincent Corvinelli}, {and}
  \bibinfo{person}{Calisto Zuzarte}.} \bibinfo{year}{2015}\natexlab{}.
\newblock \showarticletitle{Cardinality estimation using neural networks}. In
  \bibinfo{booktitle}{\emph{Proceedings of the 25th Annual International
  Conference on Computer Science and Software Engineering}}.
  \bibinfo{pages}{53--59}.
\newblock


\bibitem[\protect\citeauthoryear{Liu}{Liu}{2009}]%
        {Liu09:Learning}
\bibfield{author}{\bibinfo{person}{Tie-Yan Liu}.}
  \bibinfo{year}{2009}\natexlab{}.
\newblock \showarticletitle{Learning to rank for information retrieval}.
\newblock \bibinfo{journal}{\emph{Foundations and Trends in Information
  Retrieval}} \bibinfo{volume}{3}, \bibinfo{number}{3} (\bibinfo{year}{2009}),
  \bibinfo{pages}{225--331}.
\newblock


\bibitem[\protect\citeauthoryear{Lohman}{Lohman}{2014}]%
        {provequeryhard_lohman2014query}
\bibfield{author}{\bibinfo{person}{Guy Lohman}.}
  \bibinfo{year}{2014}\natexlab{}.
\newblock \showarticletitle{Is query optimization a “solved” problem}. In
  \bibinfo{booktitle}{\emph{Proc. Workshop on Database Query Optimization}},
  Vol.~\bibinfo{volume}{13}. Oregon Graduate Center Comp. Sci. Tech. Rep,
  \bibinfo{pages}{10}.
\newblock


\bibitem[\protect\citeauthoryear{Luce}{Luce}{1959}]%
        {Luce59:Individual}
\bibfield{author}{\bibinfo{person}{R~Duncan Luce}.}
  \bibinfo{year}{1959}\natexlab{}.
\newblock \bibinfo{title}{Individual Choice Behavior}.
\newblock
\newblock


\bibitem[\protect\citeauthoryear{Marcus, Negi, Mao, Tatbul, Alizadeh, and
  Kraska}{Marcus et~al\mbox{.}}{2022}]%
        {marcus2022bao}
\bibfield{author}{\bibinfo{person}{Ryan Marcus}, \bibinfo{person}{Parimarjan
  Negi}, \bibinfo{person}{Hongzi Mao}, \bibinfo{person}{Nesime Tatbul},
  \bibinfo{person}{Mohammad Alizadeh}, {and} \bibinfo{person}{Tim Kraska}.}
  \bibinfo{year}{2022}\natexlab{}.
\newblock \showarticletitle{Bao: Making learned query optimization practical}.
\newblock \bibinfo{journal}{\emph{ACM SIGMOD Record}} \bibinfo{volume}{51},
  \bibinfo{number}{1} (\bibinfo{year}{2022}), \bibinfo{pages}{6--13}.
\newblock


\bibitem[\protect\citeauthoryear{Marcus and Papaemmanouil}{Marcus and
  Papaemmanouil}{2018}]%
        {rejoin_marcus2018deep}
\bibfield{author}{\bibinfo{person}{Ryan Marcus} {and} \bibinfo{person}{Olga
  Papaemmanouil}.} \bibinfo{year}{2018}\natexlab{}.
\newblock \showarticletitle{Deep reinforcement learning for join order
  enumeration}. In \bibinfo{booktitle}{\emph{Proceedings of the First
  International Workshop on Exploiting Artificial Intelligence Techniques for
  Data Management}}. \bibinfo{pages}{1--4}.
\newblock


\bibitem[\protect\citeauthoryear{Marcus, Negi, Mao, Zhang, Alizadeh, Kraska,
  Papaemmanouil, and Tatbul}{Marcus et~al\mbox{.}}{2019}]%
        {neo}
\bibfield{author}{\bibinfo{person}{Ryan~C. Marcus}, \bibinfo{person}{Parimarjan
  Negi}, \bibinfo{person}{Hongzi Mao}, \bibinfo{person}{Chi Zhang},
  \bibinfo{person}{Mohammad Alizadeh}, \bibinfo{person}{Tim Kraska},
  \bibinfo{person}{Olga Papaemmanouil}, {and} \bibinfo{person}{Nesime Tatbul}.}
  \bibinfo{year}{2019}\natexlab{}.
\newblock \showarticletitle{Neo: {A} Learned Query Optimizer}.
\newblock \bibinfo{journal}{\emph{Proc. {VLDB} Endow.}} \bibinfo{volume}{12},
  \bibinfo{number}{11} (\bibinfo{year}{2019}), \bibinfo{pages}{1705--1718}.
\newblock
\urldef\tempurl%
\url{https://doi.org/10.14778/3342263.3342644}
\showDOI{\tempurl}


\bibitem[\protect\citeauthoryear{Mitchell}{Mitchell}{1980}]%
        {inductivebias_mitchell1980need}
\bibfield{author}{\bibinfo{person}{Tom~M Mitchell}.}
  \bibinfo{year}{1980}\natexlab{}.
\newblock \bibinfo{booktitle}{\emph{The need for biases in learning
  generalizations}}.
\newblock \bibinfo{publisher}{Citeseer}.
\newblock


\bibitem[\protect\citeauthoryear{Mou, Li, Zhang, Wang, and Jin}{Mou
  et~al\mbox{.}}{2016}]%
        {tcnn_mou2016convolutional}
\bibfield{author}{\bibinfo{person}{Lili Mou}, \bibinfo{person}{Ge Li},
  \bibinfo{person}{Lu Zhang}, \bibinfo{person}{Tao Wang}, {and}
  \bibinfo{person}{Zhi Jin}.} \bibinfo{year}{2016}\natexlab{}.
\newblock \showarticletitle{Convolutional neural networks over tree structures
  for programming language processing}. In \bibinfo{booktitle}{\emph{Thirtieth
  AAAI conference on artificial intelligence}}.
\newblock


\bibitem[\protect\citeauthoryear{Plackett}{Plackett}{1975}]%
        {Plackett75:Analysis}
\bibfield{author}{\bibinfo{person}{Robin~L Plackett}.}
  \bibinfo{year}{1975}\natexlab{}.
\newblock \showarticletitle{The analysis of permutations}.
\newblock \bibinfo{journal}{\emph{Journal of the Royal Statistical Society
  Series C: Applied Statistics}} \bibinfo{volume}{24}, \bibinfo{number}{2}
  (\bibinfo{year}{1975}), \bibinfo{pages}{193--202}.
\newblock


\bibitem[\protect\citeauthoryear{Poess and Floyd}{Poess and Floyd}{2000}]%
        {tpch_poess2000new}
\bibfield{author}{\bibinfo{person}{Meikel Poess} {and} \bibinfo{person}{Chris
  Floyd}.} \bibinfo{year}{2000}\natexlab{}.
\newblock \showarticletitle{New TPC benchmarks for decision support and web
  commerce}.
\newblock \bibinfo{journal}{\emph{ACM Sigmod Record}} \bibinfo{volume}{29},
  \bibinfo{number}{4} (\bibinfo{year}{2000}), \bibinfo{pages}{64--71}.
\newblock


\bibitem[\protect\citeauthoryear{Poosala and Ioannidis}{Poosala and
  Ioannidis}{1997}]%
        {cbowoassumption_poosala1997selectivity}
\bibfield{author}{\bibinfo{person}{Viswanath Poosala} {and}
  \bibinfo{person}{Yannis~E Ioannidis}.} \bibinfo{year}{1997}\natexlab{}.
\newblock \showarticletitle{Selectivity estimation without the attribute value
  independence assumption}. In \bibinfo{booktitle}{\emph{VLDB}},
  Vol.~\bibinfo{volume}{97}. \bibinfo{pages}{486--495}.
\newblock


\bibitem[\protect\citeauthoryear{Selinger, Astrahan, Chamberlin, Lorie, and
  Price}{Selinger et~al\mbox{.}}{1979}]%
        {oldwork_selinger1979access}
\bibfield{author}{\bibinfo{person}{P~Griffiths Selinger},
  \bibinfo{person}{Morton~M Astrahan}, \bibinfo{person}{Donald~D Chamberlin},
  \bibinfo{person}{Raymond~A Lorie}, {and} \bibinfo{person}{Thomas~G Price}.}
  \bibinfo{year}{1979}\natexlab{}.
\newblock \showarticletitle{Access path selection in a relational database
  management system}. In \bibinfo{booktitle}{\emph{Proceedings of the 1979 ACM
  SIGMOD international conference on Management of data}}.
  \bibinfo{pages}{23--34}.
\newblock


\bibitem[\protect\citeauthoryear{Sun and Li}{Sun and Li}{2019}]%
        {e2ecost_19_sunji}
\bibfield{author}{\bibinfo{person}{Ji Sun} {and} \bibinfo{person}{Guoliang
  Li}.} \bibinfo{year}{2019}\natexlab{}.
\newblock \showarticletitle{An End-to-End Learning-based Cost Estimator}.
\newblock \bibinfo{journal}{\emph{Proc. {VLDB} Endow.}} \bibinfo{volume}{13},
  \bibinfo{number}{3} (\bibinfo{year}{2019}), \bibinfo{pages}{307--319}.
\newblock
\urldef\tempurl%
\url{https://doi.org/10.14778/3368289.3368296}
\showDOI{\tempurl}


\bibitem[\protect\citeauthoryear{Thompson}{Thompson}{1933}]%
        {sampling_thompson1933likelihood}
\bibfield{author}{\bibinfo{person}{William~R Thompson}.}
  \bibinfo{year}{1933}\natexlab{}.
\newblock \showarticletitle{On the likelihood that one unknown probability
  exceeds another in view of the evidence of two samples}.
\newblock \bibinfo{journal}{\emph{Biometrika}} \bibinfo{volume}{25},
  \bibinfo{number}{3-4} (\bibinfo{year}{1933}), \bibinfo{pages}{285--294}.
\newblock


\bibitem[\protect\citeauthoryear{Trummer, Wang, Wei, Maram, Moseley, Jo,
  Antonakakis, and Rayabhari}{Trummer et~al\mbox{.}}{2021}]%
        {trummer2021skinnerdb}
\bibfield{author}{\bibinfo{person}{Immanuel Trummer}, \bibinfo{person}{Junxiong
  Wang}, \bibinfo{person}{Ziyun Wei}, \bibinfo{person}{Deepak Maram},
  \bibinfo{person}{Samuel Moseley}, \bibinfo{person}{Saehan Jo},
  \bibinfo{person}{Joseph Antonakakis}, {and} \bibinfo{person}{Ankush
  Rayabhari}.} \bibinfo{year}{2021}\natexlab{}.
\newblock \showarticletitle{Skinnerdb: Regret-bounded query evaluation via
  reinforcement learning}.
\newblock \bibinfo{journal}{\emph{ACM Transactions on Database Systems (TODS)}}
  \bibinfo{volume}{46}, \bibinfo{number}{3} (\bibinfo{year}{2021}),
  \bibinfo{pages}{1--45}.
\newblock


\bibitem[\protect\citeauthoryear{Tsesmelis and Simitsis}{Tsesmelis and
  Simitsis}{2022}]%
        {aidbsurvey_tsesmelis2022database}
\bibfield{author}{\bibinfo{person}{Dimitris Tsesmelis} {and}
  \bibinfo{person}{Alkis Simitsis}.} \bibinfo{year}{2022}\natexlab{}.
\newblock \showarticletitle{Database Optimizers in the Era of Learning}. In
  \bibinfo{booktitle}{\emph{2022 IEEE 38th International Conference on Data
  Engineering (ICDE)}}. IEEE, \bibinfo{pages}{3213--3216}.
\newblock


\bibitem[\protect\citeauthoryear{Ventura, Kaoudi, Quian{\'e}-Ruiz, and
  Markl}{Ventura et~al\mbox{.}}{2021}]%
        {ventura2021expand}
\bibfield{author}{\bibinfo{person}{Francesco Ventura}, \bibinfo{person}{Zoi
  Kaoudi}, \bibinfo{person}{Jorge~Arnulfo Quian{\'e}-Ruiz}, {and}
  \bibinfo{person}{Volker Markl}.} \bibinfo{year}{2021}\natexlab{}.
\newblock \showarticletitle{Expand your training limits! generating training
  data for ml-based data management}. In \bibinfo{booktitle}{\emph{Proceedings
  of the 2021 International Conference on Management of Data}}.
  \bibinfo{pages}{1865--1878}.
\newblock


\bibitem[\protect\citeauthoryear{Wu, Chi, Zhu, Tatemura, Hacig{\"u}m{\"u}s, and
  Naughton}{Wu et~al\mbox{.}}{2013}]%
        {wu2013predicting}
\bibfield{author}{\bibinfo{person}{Wentao Wu}, \bibinfo{person}{Yun Chi},
  \bibinfo{person}{Shenghuo Zhu}, \bibinfo{person}{Junichi Tatemura},
  \bibinfo{person}{Hakan Hacig{\"u}m{\"u}s}, {and} \bibinfo{person}{Jeffrey~F
  Naughton}.} \bibinfo{year}{2013}\natexlab{}.
\newblock \showarticletitle{Predicting query execution time: Are optimizer cost
  models really unusable?}. In \bibinfo{booktitle}{\emph{2013 IEEE 29th
  International Conference on Data Engineering (ICDE)}}. IEEE,
  \bibinfo{pages}{1081--1092}.
\newblock


\bibitem[\protect\citeauthoryear{Xia, Liu, Wang, Zhang, and Li}{Xia
  et~al\mbox{.}}{2008}]%
        {listmle_xia2008listwise}
\bibfield{author}{\bibinfo{person}{Fen Xia}, \bibinfo{person}{Tie-Yan Liu},
  \bibinfo{person}{Jue Wang}, \bibinfo{person}{Wensheng Zhang}, {and}
  \bibinfo{person}{Hang Li}.} \bibinfo{year}{2008}\natexlab{}.
\newblock \showarticletitle{Listwise approach to learning to rank: theory and
  algorithm}. In \bibinfo{booktitle}{\emph{Proceedings of the 25th
  international conference on Machine learning}}. \bibinfo{pages}{1192--1199}.
\newblock


\bibitem[\protect\citeauthoryear{Xu, Wang, Chen, and Li}{Xu
  et~al\mbox{.}}{2015}]%
        {leakyrelu_xu2015empirical}
\bibfield{author}{\bibinfo{person}{Bing Xu}, \bibinfo{person}{Naiyan Wang},
  \bibinfo{person}{Tianqi Chen}, {and} \bibinfo{person}{Mu Li}.}
  \bibinfo{year}{2015}\natexlab{}.
\newblock \showarticletitle{Empirical evaluation of rectified activations in
  convolutional network}.
\newblock \bibinfo{journal}{\emph{arXiv preprint arXiv:1505.00853}}
  (\bibinfo{year}{2015}).
\newblock


\bibitem[\protect\citeauthoryear{Yang, Chiang, Luan, Mittal, Luo, and
  Stoica}{Yang et~al\mbox{.}}{2022}]%
        {balsa_SIGMOD22}
\bibfield{author}{\bibinfo{person}{Zongheng Yang}, \bibinfo{person}{Wei{-}Lin
  Chiang}, \bibinfo{person}{Sifei Luan}, \bibinfo{person}{Gautam Mittal},
  \bibinfo{person}{Michael Luo}, {and} \bibinfo{person}{Ion Stoica}.}
  \bibinfo{year}{2022}\natexlab{}.
\newblock \showarticletitle{Balsa: Learning a Query Optimizer Without Expert
  Demonstrations}. In \bibinfo{booktitle}{\emph{{SIGMOD} '22: International
  Conference on Management of Data, Philadelphia, PA, USA, June 12 - 17,
  2022}}, \bibfield{editor}{\bibinfo{person}{Zachary Ives},
  \bibinfo{person}{Angela Bonifati}, {and} \bibinfo{person}{Amr~El Abbadi}}
  (Eds.). \bibinfo{publisher}{{ACM}}, \bibinfo{pages}{931--944}.
\newblock
\urldef\tempurl%
\url{https://doi.org/10.1145/3514221.3517885}
\showDOI{\tempurl}


\bibitem[\protect\citeauthoryear{Yang, Kamsetty, Luan, Liang, Duan, Chen, and
  Stoica}{Yang et~al\mbox{.}}{2020}]%
        {yang2020neurocard}
\bibfield{author}{\bibinfo{person}{Zongheng Yang}, \bibinfo{person}{Amog
  Kamsetty}, \bibinfo{person}{Sifei Luan}, \bibinfo{person}{Eric Liang},
  \bibinfo{person}{Yan Duan}, \bibinfo{person}{Xi Chen}, {and}
  \bibinfo{person}{Ion Stoica}.} \bibinfo{year}{2020}\natexlab{}.
\newblock \showarticletitle{NeuroCard: one cardinality estimator for all
  tables}.
\newblock \bibinfo{journal}{\emph{Proceedings of the VLDB Endowment}}
  \bibinfo{volume}{14}, \bibinfo{number}{1} (\bibinfo{year}{2020}),
  \bibinfo{pages}{61--73}.
\newblock


\bibitem[\protect\citeauthoryear{Yang, Liang, Kamsetty, Wu, Duan, Chen, Abbeel,
  Hellerstein, Krishnan, and Stoica}{Yang et~al\mbox{.}}{2019}]%
        {naru_yang2019deep}
\bibfield{author}{\bibinfo{person}{Zongheng Yang}, \bibinfo{person}{Eric
  Liang}, \bibinfo{person}{Amog Kamsetty}, \bibinfo{person}{Chenggang Wu},
  \bibinfo{person}{Yan Duan}, \bibinfo{person}{Xi Chen},
  \bibinfo{person}{Pieter Abbeel}, \bibinfo{person}{Joseph~M Hellerstein},
  \bibinfo{person}{Sanjay Krishnan}, {and} \bibinfo{person}{Ion Stoica}.}
  \bibinfo{year}{2019}\natexlab{}.
\newblock \showarticletitle{Deep unsupervised cardinality estimation}.
\newblock \bibinfo{journal}{\emph{arXiv preprint arXiv:1905.04278}}
  (\bibinfo{year}{2019}).
\newblock


\bibitem[\protect\citeauthoryear{Yu, Piryani, Jatowt, Inagaki, Joho, and
  Kim}{Yu et~al\mbox{.}}{2023}]%
        {yu2023depth}
\bibfield{author}{\bibinfo{person}{Hai-Tao Yu}, \bibinfo{person}{Rajesh
  Piryani}, \bibinfo{person}{Adam Jatowt}, \bibinfo{person}{Ryo Inagaki},
  \bibinfo{person}{Hideo Joho}, {and} \bibinfo{person}{Kyoung-Sook Kim}.}
  \bibinfo{year}{2023}\natexlab{}.
\newblock \showarticletitle{An in-depth study on adversarial learning-to-rank}.
\newblock \bibinfo{journal}{\emph{Information Retrieval Journal}}
  \bibinfo{volume}{26}, \bibinfo{number}{1} (\bibinfo{year}{2023}),
  \bibinfo{pages}{1}.
\newblock


\bibitem[\protect\citeauthoryear{Zhang, Yin, Zhu, and Zhang}{Zhang
  et~al\mbox{.}}{2018}]%
        {representationLearningSurvey_zhang2018network}
\bibfield{author}{\bibinfo{person}{Daokun Zhang}, \bibinfo{person}{Jie Yin},
  \bibinfo{person}{Xingquan Zhu}, {and} \bibinfo{person}{Chengqi Zhang}.}
  \bibinfo{year}{2018}\natexlab{}.
\newblock \showarticletitle{Network representation learning: A survey}.
\newblock \bibinfo{journal}{\emph{IEEE transactions on Big Data}}
  \bibinfo{volume}{6}, \bibinfo{number}{1} (\bibinfo{year}{2018}),
  \bibinfo{pages}{3--28}.
\newblock


\bibitem[\protect\citeauthoryear{Zhang}{Zhang}{2020}]%
        {alphajoin_zhang2020alphajoin}
\bibfield{author}{\bibinfo{person}{Ji Zhang}.} \bibinfo{year}{2020}\natexlab{}.
\newblock \showarticletitle{AlphaJoin: Join Order Selection {\`a} la AlphaGo}.
  In \bibinfo{booktitle}{\emph{PVLDB-PhD}}.
\newblock


\bibitem[\protect\citeauthoryear{Zhao, Liu, and Xia}{Zhao
  et~al\mbox{.}}{2022}]%
        {Zhao22:Learning}
\bibfield{author}{\bibinfo{person}{Zhibing Zhao}, \bibinfo{person}{Ao Liu},
  {and} \bibinfo{person}{Lirong Xia}.} \bibinfo{year}{2022}\natexlab{}.
\newblock \showarticletitle{Learning mixtures of random utility models with
  features from incomplete preferences}.
\newblock \bibinfo{journal}{\emph{arXiv preprint arXiv:2006.03869}}
  (\bibinfo{year}{2022}).
\newblock


\bibitem[\protect\citeauthoryear{Zhao and Xia}{Zhao and Xia}{2018}]%
        {Zhao18:Composite}
\bibfield{author}{\bibinfo{person}{Zhibing Zhao} {and} \bibinfo{person}{Lirong
  Xia}.} \bibinfo{year}{2018}\natexlab{}.
\newblock \showarticletitle{Composite marginal likelihood methods for random
  utility models}. In \bibinfo{booktitle}{\emph{International Conference on
  Machine Learning}}. PMLR, \bibinfo{pages}{5922--5931}.
\newblock


\end{thebibliography}

\end{document}